\documentclass[12pt]{iopart}
\pdfoutput=1
\usepackage{graphicx}
\expandafter\let\csname equation*\endcsname\relax
\expandafter\let\csname endequation*\endcsname\relax
\usepackage{amsmath}
\usepackage{amssymb}
\usepackage{bm}
\usepackage{cancel}
\usepackage[normalem]{ulem}
\usepackage{subfig}
\usepackage{placeins}
\usepackage{color}
\usepackage{iopams}
\usepackage{cite}

\usepackage{hyperref}
\newcommand{\tder}[1]{\partial_{t}{#1}}
\newcommand{\tauder}[1]{\partial_{\tau}{#1}}
\newcommand{\xder}[1]{\partial_{x}{#1}}
\newcommand{\sder}[1]{\partial_{s}{#1}}

\newcommand{\sigdertot}[1]{\frac{d}{d\sigma}{#1}}

\newcommand{\vv}{\bm{v}}
\newcommand{\HCS}{\text{\tiny HCS}}

\newcommand{\eq}{\text{\tiny eq}}

\newcommand{\av}{\text{av}}
\newcommand{\llangle}{\left\langle}
\newcommand{\rrangle}{\right\rangle}

\begin{document}


\title{Lattice models for granular-like velocity fields: Finite-size effects}


\author{C.~A.~Plata$^{1}$, A.~ Manacorda$^{2}$, A.~ Lasanta$^{3,4}$,
 A.~ Puglisi$^{3}$, and A. Prados$^{1}$ }

\address{
$^{1}$F\'{\i}sica Te\'{o}rica, Universidad de Sevilla,
Apartado de Correos 1065, E-41080 Seville, Spain}
\address{
$^{2}$Dipartimento di Fisica, Sapienza Universit\`a di Roma,
p.le A. Moro 2, 00185 Roma, Italy}
\address{
$^{3}$CNR-ISC and Dipartimento di Fisica, Sapienza
Universit\`a di Roma, p.le A. Moro 2, 00185 Roma, Italy}
\address{
$^{4}$ Departamento de F\'{\i}sica. Universidad de Extremadura,        
E-06071 Badajoz, Spain}

\begin{abstract}
  Long-range spatial correlations in the velocity and energy fields of
  a granular fluid are discussed in the framework of a 1d lattice
  model. The dynamics of the velocity field occurs through
  nearest-neighbour inelastic collisions that conserve momentum but
  dissipate energy. A set of equations for the fluctuating
  hydrodynamics of the velocity and energy mesoscopic fields give a
  first approximation for (i) the velocity structure factor and (ii)
  the finite-size correction to the Haff law, both in the homogeneous
  cooling regime. At a more refined level, we have derived the
  equations for the two-site velocity correlations
  and the total energy fluctuations. First, we seek
    a perturbative solution thereof, in powers of the inverse of
    system size. On the one hand, when scaled with the granular
    temperature, the velocity correlations tend to a stationary value
    in the long time limit. On the other hand, the scaled standard
    deviation of the total energy diverges, that is, the system shows
    multiscaling. Second, we find an exact solution for the velocity
    correlations in terms of the spectrum of eigenvalues of a certain
  matrix. The results of numerical simulations of the microscopic
  model confirm our theoretical results, including the above
    described multiscaling phenomenon.
\end{abstract}
\section{Introduction}
\label{sec:intro}

Granular materials are systems that comprise a ``large'' number of
macroscopic particles (grains), the collisions between which are
inelastic~\cite{jaeger96}. When the grains inside a container are
rapidly shaken, the granular fluid regime appears. Moreover, if the
particles move freely between their binary instantaneous collisions,
we have a ``granular gas'' ~\cite{puglbook}.  This is usually so when
the peak acceleration is many times the gravity and the packing
fraction is below $\sim\! 1\%$. Interestingly, kinetic theory can be
extended to the granular case by writing an inelastic version of the
Boltzmann equation~\cite{BP04}, which takes into account that energy
is no longer conserved in collisions.

The number of particles in a granular system is much smaller than
Avogadro's number, and then fluctuations are expected to be
more important than those in molecular systems.  A promising approach
is to study intermediate coarse-graining schemes, such as mesoscospic
fluctuating hydrodynamics, to account for the fluctuations of ``slow
variables'' in the system, e.g. those associated to conserved or
quasi-conserved quantities such as momentum or energy. Regrettably,
there is not a generalisation of the theory of mesoscopic fluctuations
valid at equilibrium~\cite{eins,onsager,landau} to the non-equilibrium
domain.  Nevertheless, some progress have been recently made to derive
a consistent fluctuating hydrodynamic picture from the inelastic
Boltzmann equation ~\cite{BMG09}. Here we choose a different
perspective, looking for a simplified model that contains the
essential ingredients of granular fluids but reduces the complexity of
the task and promises a transparent interpretation of the results. In
particular we aim to elucidate the ``perturbative'' nature of the
continuum limit and calculate the corrections to
it~\cite{spohn80}. Such corrections give interesting information about
the structure - in space and time - of the correlated granular
fluctuations and reveal new phenomena peculiar of inelastic
collisions.

Lattice models are a useful tool in non-equilibrium statistical
mechanics. Their simplicity makes it possible to identify the main
elements to describe the experimentally relevant behaviour in many
different physical situations. More specifically, lattice models have
been of paramount importance for understanding rigorously the
conditions needed to have a hydrodynamic description, both at the
average ~\cite{kl,kmp} and fluctuating \cite{Bertini}
levels. Recently, fluctuating hydrodynamics has been employed to
derive the large deviation function in the context of
energy-conserving models \cite{Pablo,Pablo2,Pablo3,HyK11} and even in
energy-dissipating models~\cite{PLyH12a,PLyH11a,PLyH13}. Both in the
conservative and non-conservative case, momentum conservation has not
been taken into account. This shortcoming may be relevant, since it is
known that momentum conservation is linked to the appearance of
long-ranged correlations in out-of-equilibrium
systems~\cite{grinstein,correlations}.

Very recently, a lattice model for the velocity field with momentum
conservation has been put forward~\cite{noi}. In a previous paper
\cite{first}, we derived both the average and fluctuating hydrodynamic
pictures, and looked into some relevant physical situations,
such as the Homogeneous Cooling or Uniform Shear Flow states. In the
present paper we focus on the spatial long-range velocity correlations
in the Homogeneous Cooling State, which can be partly explained by
fluctuating hydrodynamics but require a more refined treatment to be
fully investigated. In our study of velocity and energy spatial
correlations two original phenomena emerge: a correction to the Haff's
law due to spatial correlations and a multiscaling phenomenon where
the total energy fluctuations do not scale with the temperature.

We briefly summarise the organisation of the paper. In
section~\ref{sec:model}, we introduce the microscopic lattice model,
its Master Equation and the fluctuating hydrodynamic equations which
have been first presented in~\cite{noi} and discussed in detail
in~\cite{first}. Section~\ref{sec:meso-fluc-th} is devoted
to the Homogeneous Cooling State (HCS). Therein, we derive
the shape of velocity and energy spatial correlations from fluctuating
hydrodynamics, and also compute the first order finite size correction
to the Haff law. The microscopic equations for the velocity
correlations are obtained in
  section~\ref{sec:beyond_mol_chaos}. By going to the continuum limit
  in them, the energy decay in the HCS is computed in a more precise
form that is valid for longer times. In section~\ref{sec:HCS-exact},
we discuss an exact scheme for the resolution of the microscopic
correlation equations, before taking the continuum limit.
The presence of multiscaling for the energy fluctuations is presented
in section~\ref{sec:total_energy}, through both numerical
results and an approximate theory based on clustering the
  three-particle correlations.  Conclusions and perspectives are
presented in section~\ref{sec:concl}. The appendices deal
  with some technical details that are omitted in the main text.

\section{Revision of the model and previous results}
\label{sec:model}

In this section, we briefly revise the main aspects of the model
introduced in \cite{noi}, focusing on its continuum,
hydrodynamic-like, limit. A more detailed description of the model can
be found in \cite{first}.

\subsection{Dynamics}

Let us consider a 1d lattice with $N$ sites. First, we define the
dynamics in discrete time: after the $p$-th step of the dynamics, the
particle at the $l$-th site has a velocity $v_{l,p} \in$
$\mathcal{R}$. The configuration for the system at time $p$ is denoted
as $\vv_p \equiv \{v_{1,p},...,v_{N,p}\}$. One individual trajectory
of the (Markovian) stochastic process is built in the following way:
the configuration of the system changes from time $p$ to time $p+1$
because a pair of nearest neighbours $\left( l,l+1 \right)$ is chosen
at random and collides inelastically, that is $\vv_{p+1} = \hat{b}_l
\vv_p$ where the operator $\hat{b}_{l}$ transforms the pre-collisional
velocities $(v_{l,p},v_{l+1,p})$ into the post-collisional ones
$(v_{l,p+1},v_{l+1,p+1})$ and leaves all other sites unaltered. The
post-collisional velocities are given by
\begin{subequations}\label{coll_rule}
\begin{eqnarray}
v_{l,p+1} &=& v_{l,p}-(1+\alpha)\Delta_{l,p}/2 \\
v_{l+1,p+1} &=& v_{l+1,p}+(1+\alpha)\Delta_{l,p}/2,
\end{eqnarray}
\end{subequations}
with $\Delta_{l,p}=v_{l,p}-v_{l+1,p}$ and $0<\alpha\leq 1$. In the
following we use a notation such that the evolution operator
$\hat{b}_{l}$ acts naturally on {\em observables}, e.g. $\hat{b}_l
v_{l,p} = v_{l,p+1}$. Momentum is always conserved,
$(\hat{b}_{l}-1)(v_{l,p}+v_{l+1,p})=0$, whereas energy, if $\alpha
\neq 1$, is not:
$(\hat{b}_{l}-1)(v^{2}_{l,p}+v^{2}_{l+1,p})=(\alpha^2-1)\Delta_{l,p}^{2}/2<0$. The
collision rule \eqref{coll_rule}, which corresponds to the simplest
one used in granular fluids \cite{PyL01}, is valid for bulk sites and
must be complemented with suitable boundary conditions.

The evolution equation for the velocities can be cast in the form
\begin{equation}\label{eq:mom}
v_{l,p+1}-v_{l,p}=-j_{l,p}+j_{l-1,p}, \quad j_{l,p}=\frac{1+\alpha}{2}\Delta_{l,p}\delta_{y_p,l}.
\end{equation}
This is nothing but a discrete continuity equation, in which $j_{l,p}$
is the momentum current from site $l$ to site $l+1$ at time $p$.  In
\eqref{eq:mom}, $\delta_{y_{p},l}$ is Kronecker's delta and $y_{p}$
is a homogeneously distributed random integer in $[1,L]$, where $L$ is
the number of possible colliding pairs. For periodic
boundary conditions, $L=N$, whereas for thermostatted boundaries
$L=N+1$ \cite{first}. 

We have only kinetic energy, which is
${\cal K}_{p}=\sum_{l=1}^{N} e_{l,p}$ at time $p$, where
$e_{l,p}=v_{l,p}^{2}$. By squaring (\ref{eq:mom}), the evolution
equation for the energy at site $l$ is obtained
\begin{eqnarray}\label{eq:en}
 e_{l,p+1}-e_{l,p}=-J_{l,p}+J_{l-1,p}+d_{l,p}.
\end{eqnarray}
Apart from the ``flux'' term $-J_{l,p}+J_{l-1,p}$, we have a sink term
$d_{l,p}$ that stems from the inelasticity of collisions. The energy
current $J_{l,p}$ from site $l$ to site $l+1$ and energy dissipation $d_{l,p}$
at site $l$ are 
\begin{equation}\label{micro-ener-dis}
J_{l,p}=(v_{l,p}+v_{l+1,p})j_{l,p}, \quad
d_{l,p}= \frac{\alpha^2-1}{4}\left[\delta_{y_p,l}\Delta_{l,p}^2+\delta_{y_p,l-1}\Delta_{l-1,p}^2 \right]<0,
\end{equation}
respectively.

The above stochastic dynamics generates the trajectories that
correspond to the Markov process described by the master equation in
continuous time \cite{first}
\begin{equation} \label{eq:cma2}
\partial_\tau P_N(\vv,\tau|\vv_0,\tau_0)=\omega \sum_{l=1}^L \left[ \frac{P_{N}(\hat{b}_l^{-1} \vv,\tau|\vv_0,\tau_0)}{\alpha}  -  P_N(\vv,\tau|\vv_0,\tau_0) \right],
\end{equation}
for the conditional probability density
  $P_{N}(\vv,\tau|\vv_0,\tau_0)$ of finding the system in state $\vv$
  at time $\tau$ provided it was in state $\vv_{0}$ at time
  $\tau_{0}$.  Above, $\omega$ is a constant frequency that
determines the time scale and the operator $\hat{b}_l^{-1}$ is the
inverse of $\hat{b}_{l}$. Thus, $\hat{b}_l^{-1}$ changes the
post-collisional velocities into the pre-collisional ones when the
colliding pair is $(l,l+1)$.  Moreover, the increment $\delta\tau_{p}$
of the continuous time $\tau$ at the $p$-th step of the dynamics over
each trajectory is given by $\delta\tau_{p}=-(L\omega)^{-1}\ln x$, in
which $x$ is a stochastic variable homogeneously distributed in the
interval $(0,1)$.

\subsection{Hydrodynamic equations}\label{sec:hydro-eq}

In the large system size limit as $L\to\infty$, a continuum limit may
be introduced by considering that the average velocity
$u_{l,p}=\langle v_{l,p}\rangle$ and energy
$E_{l,p}=\langle v_{l,p}^{2}\rangle$ are smooth functions of space and
time. Of course, the local temperature $T_{l,p}=E_{l,p}-u_{l,p}^{2}$
is also smooth.  Specifically, we introduce hydrodynamic continuous
space and time variables $x=l/L$ and $t=\omega\tau/L^{2}$,
respectively.

In this continuum limit, the balance equations for the average
velocity $u(x,t)$ and energy $E(x,t)=u^{2}(x,t)+T(x,t)$ read
\begin{subequations}\label{eq:av-u-E}
\begin{align}
\partial_t u(x,t) &= -\partial_x j_{\av}(x,t) , \\
\partial_t E(x,t) &=-\partial_x J_{\av}(x,t) + d_{\av}(x,t).
\end{align}
\end{subequations}
Therein, the average momentum and energy currents,  $j_{\av}(x,t)$ and 
$J_{\av}(x,t)$, respectively,   are given by
\begin{equation}\label{eq:av-j-J}
  j_{\av}(x,t)=-\partial_{x} u(x,t), \quad
  J_{\av}(x,t)=-\partial_{x} E(x,t),
\end{equation}
and the dissipation field $d_{\av}(x,t)$ is
\begin{equation}\label{eq:av-d-nu}
  d_{\av}(x,t)=-\nu T, \quad
  \nu=(1-\alpha^{2})L^{2}.
\end{equation}
We have introduced the macroscopic dissipation coefficient $\nu$,
which is the relevant parameter in the hydrodynamic space and time
scales \cite{first}. It is straightforward to combine
\eqref{eq:av-u-E}, \eqref{eq:av-j-J} and \eqref{eq:av-d-nu} to write
closed equations for the hydrodynamic fields: average velocity and
temperature, 
\begin{subequations}\label{eq:hydroMM}
\begin{align}
\partial_{t}u(x,t)&=\partial_{xx} u(x,t), \label{eq:hydroMMu} \\
\partial_{t}T(x,t)&=-\nu T(x,t)+ \partial_{xx}T(x,t)+2\left[\partial_{x}u(x,t)\right]^2 . \label{eq:hydroMMT}
\end{align}
\end{subequations}
These equations must be solved submitted to suitable boundary
conditions, which depend on the physical state under scrutiny.



\subsection{Fluctuating hydrodynamics}\label{sec:fluc-hydro}

The balance equations~\eqref{eq:av-u-E} may also be written
at the fluctuating level of description, by considering that $v(x,t)$
and $e(x,t)$ are fluctuating quantities, whose averages are $u(x,t)$
and $E(x,t)$. In this way, fluctuating balance equations are
written for both $v(x,t)$ and $e(x,t)$, which are the continuum limit
versions of the microscopic balance equations \eqref{eq:mom} and
\eqref{eq:en}, namely
\begin{subequations}\label{eq:fluct-hydro-v-e}
\begin{eqnarray}
\tder v(x,t)&=&-\partial_{x} j(x,t), \qquad\qquad\;\;\,
j(x,t)=-\partial_{x}v(x,t)+\xi^{(j)}(x,t), \\
\tder e(x,t)&=&-\partial_{x}J(x,t)+d(x,t), \;\; J(x,t)=-\partial_{x}e(x,t)+\xi^{(J)}(x,t).
\end{eqnarray}
\end{subequations}
In the equations above, $(j,J)$ are the fluctuating currents for
momentum and energy, and $(\xi^{(j)},\xi^{(J)})$ are their
corresponding noises.  These noises have been shown to be Gaussian and white~\cite{first}. The
amplitudes of their correlations $\langle
\xi^{(\gamma)}\xi^{(\gamma')}\rangle$ can be cast in matrix form, 
\begin{equation}
  \label{eq:noise-corr}
  \langle
  \xi^{(\gamma)}(x,t)\xi^{(\gamma')}(x',t')\rangle=L^{-1}
  \bm{\Xi}^{(\gamma\gamma')}\delta(x-x')\delta(t-t'),
\end{equation}
 where $(1,2)$ for $\gamma$ or $\gamma'$ correspond to
$(j,J)$, and have been computed
  within the local equilibrium approximation in~\cite{first}, with the result
\begin{equation}\label{eq:corr-matrix}
   \bm{\Xi}=2T(x,t)
\begin{pmatrix}
    1 & & 2u(x,t) \\
   2u(x,t) & & 2[T(x,t)+2 u^2(x,t)]
 \end{pmatrix}.
\end{equation}
The average velocity $u(x,t)$ and the temperature $T(x,t)$ must be
calculated in the state corresponding to the physical situation of
interest.


Finally, the dissipation field $d(x,t)$ is given by
\begin{equation}
  \label{eq:fluct-d}
  d(x,t)=-\nu \theta(x,t)=-\nu \left[e(x,t)-v_{R}^{2}(x,t)\right],
\end{equation}
where $v_{R}^{2}$ is the regular part of $v^{2}$, defined as
\begin{equation}\label{vR-main-text}
  v_{R}^{2}(x,t)\equiv v^{2}(x,t)-L^{-1}\theta(x,t)
  \lim_{\Delta x\to 0}\delta(\Delta x).
\end{equation}
This regular part of the velocity field has the property
$\langle v_{R} ^{2}(x,t)\rangle =u^{2}(x,t)$, as shown
in~\ref{app-a}. Equation~\eqref{eq:fluct-d} tells us that the
fluctuations of the dissipation field are enslaved to those of the
fluctuating temperature field $\theta(x,t)$.  This is so since the
dissipation noise $\xi^{(d)}$ is subdominant as compared to the
current noises, because it scales as $L^{-3}$ instead of as $L^{-1}$,
as proven in \cite{first}.

\section{Mesoscopic Fluctuation Theory}
\label{sec:meso-fluc-th}

To be concrete, we focus on fluctuations around the HCS, which have
already been analysed in the literature for a hard-sphere system
described by the Boltzmann equation close to the shear instability
\cite{BDGyM06}. To do so, it is useful to go to Fourier
  space by considering that all the fields are written as
\begin{equation}
  \label{eq:Fourier-def}
  y(x,t)=\sum_{n} y_{n}(t) e^{i k_{n}x}, \quad
  y_{n}(t)=\int_{0}^{1} dx\, y(x,t)   e^{-i k_{n}x}, \; k_{n}=2n\pi.
\end{equation}


\subsection{Velocity fluctuations}

The equation for the fluctuating velocity is closed,
\begin{equation}
  \label{eq:fluc-veloc}
 \tder{v}=\partial_{xx}v-\partial_{x}\xi^{(j)},
\end{equation}
and going to Fourier space,
\begin{equation}
  \label{eq:fluc-veloc-Fourier}
  \tder{v}_{n}=-k_{n}^{2}v_{n}-i k\xi_{n}^{(j)}.
\end{equation}
The long time behaviour of the solution to
\eqref{eq:fluc-veloc-Fourier} is readily obtained by taking the
initial time $t_{0}$ to $-\infty$, and then
\begin{equation}
  \label{eq:fluc-veloc-longtime}
  v_{n}(t)=-ik_{n} \int_{-\infty}^{t} ds\, e^{-k_{n}^{2}(t-s)} \xi_{n}^{(j)}(s).
\end{equation}
Now, we compute the equal-time velocity correlation in Fourier space,
\begin{equation}
  \label{eq:fluc-veloc-corr-Fourier}
  \langle v_{n}(t) v_{n'}(t)\rangle_{\HCS}=-k^{2}\int_{-\infty}^{t}
  \!\!\! ds \,
  e^{-k^{2}(t-s)}\int_{-\infty}^{t}\!\!\! ds' e^{-k^{2}(t-s')} \langle \xi_{n}^{(j)}(s)\xi_{n'}^{(j)}(s')\rangle_{\HCS}.
\end{equation}
Making use of the time dependence of the temperature in the HCS, i.e
the Haff law, we get to the lowest order
\begin{equation}
  \label{eq:veloc-corr-final}
  \langle v_{n}(t) v_{n'}(t)\rangle_{\HCS}=\frac{T_{\HCS}(t)}{L}
  \frac{2k_{n}^{2}}{2k_{n}^{2}-\nu}\delta_{n,-n'}=\frac{T_{\HCS}(t)}{L} \left(1+\frac{\nu}{2k_{n}^{2}-\nu}\right)\delta_{n,-n'},
\end{equation}
provided that $2k_{n}^{2}-\nu>0$. Thus, these correlations are valid
for all $n$ when $\nu<\nu_{c}=8\pi^{2}$, since at
$\nu=\nu_{c}$ we have that $\langle v_{1}(t)v_{-1}(t)\rangle$
diverges. 

The above correlations allow us to calculate the spatial integral of
$v^{2}(x,t)$. At the fluctuating level, we have that
\begin{equation}
  \label{eq:parseval}
  \int_{0}^{1} dx \, v^{2}(x,t)=\sum_{n=-\infty}^{+\infty} v_{n}(t) v_{-n}(t),
\end{equation}
which is Parseval's theorem for the Fourier transform. By taking
averages, we readily see that $v^{2}$ has a singular contribution,
because the sum of the correlations $\langle v_{n}(t)v_{-n}(t)\rangle$~diverges. This
stems from the $\delta(0)$ contribution in \eqref{vR-main-text}, the
average value of which in the HCS is
\begin{equation}
  \label{eq:sing-part-HCS}
  \langle L^{-1}\theta(x,t)\lim_{\Delta x\to 0}\delta(\Delta x)\rangle=L^{-1}T_{\HCS}(t)\sum_{n}1,
\end{equation}
since $\delta(x-x')=\sum_{n }\exp[i k_{n}(x-x')]$. Therefore,
\begin{subequations}\label{eq:reg-psi-HCS}
\begin{eqnarray}
  \label{eq:reg-HCS} 
\hspace{12mm}  \int_{0}^{1}dx\, \langle
v_R^{2}(x,t)\rangle &=&
\frac{T_{\HCS}(t)}{L}\, \psi_{\HCS}, \\
\label{eq:psi-HCS} 
 \psi_{\HCS}(\nu)\equiv\sum_{n}
\frac{\nu}{2k_{n}^{2}-\nu}&=&-\frac{\sqrt{\nu}}{2\sqrt{2}}
\cot\left({\frac{\sqrt{\nu}}{2\sqrt{2}}}\right).
\end{eqnarray}
\end{subequations}
Of course, the spatial integral of the regular part has a finite
value. The shear instability of the HCS is clearly observed within the
framework of the fluctuating hydrodynamic description: at
$\nu=\nu_{c}=8\pi^{2}$, we have that
\begin{equation}
  \label{eq:instability}
  \lim_{\nu\to\nu_{c}}\psi_{\HCS}(\nu)=\infty,
\end{equation}
and the spatial integral of $v_{R}^{2}$ diverges. In particular, it is
$\langle v_{1}(t)v_{-1}(t) \rangle$ that diverges, as readily seen from
\eqref{eq:veloc-corr-final} and already said above.

\subsection{Effect of velocity fluctuations on the decay of average total energy}

Here, we consider the fluctuations of the total energy per particle, defined by
\begin{equation}
  \label{eq:total-energy}
  e(t)=\int_{0}^{1} dx\, e(x,t).
\end{equation}
At the mesoscopic fluctuating level, we have that
\begin{equation}
  \label{eq:total-energy-evolution}
  \frac{d}{dt}e(t)=\int_{0}^{1}dx \, d(x,t)=-\nu\, e(t) +\nu\!\int_{0}^{1}dx\,v_{R}^{2}(x,t),
\end{equation}
consistently with \eqref{eq:fluct-hydro-v-e} and
\eqref{eq:fluct-d}.

We introduce a rescaled dimensionless total energy by 
\begin{equation}\label{meso-sc-e}
 \tilde{e}(t)=\frac{e(t)}{T_{\HCS}(t)},
\end{equation}
which verifies the evolution equation
\begin{equation}
\label{eq:total-energy-sc-evol}
\frac{d}{dt}\tilde{e}(t)=\nu\!\int_{0}^{1}dx\,\tilde{v}_{R}^{2}(x,t),
\end{equation}
in which $\tilde{v}_{R}^{2}(x,t)=v_{R}^{2}(x,t)/T_{\HCS}(t)$. Now, we take averages and make use of \eqref{eq:reg-psi-HCS} to write 
\begin{equation}
  \label{eq:total-energy-sc-evol-average}
  \frac{d}{dt} \tilde{E}(t)=\psi_{\HCS}\frac{\nu}{L},
\end{equation}
which has to be integrated with the initial condition
  $\tilde{E}(0)=1$. We have omitted the $\nu$-dependence of
$\psi_{\HCS}$ in order not to clutter our formulae. Therefore, up to
order of $L^{-1}$, we have
\begin{equation}
  \label{eq:total-energy-sc-sol}
  \tilde{E}(t)=1+\delta\tilde{E}(t), \quad
  \delta{\tilde{E}}(t)=\psi_{\HCS} \frac{\nu\,t}{L}.
\end{equation}
which is expected to be valid as long as $\nu\psi_{\HCS} t/L\ll 1$.

\begin{figure}[!h]
    \centering
    \includegraphics[width=0.7\textwidth]{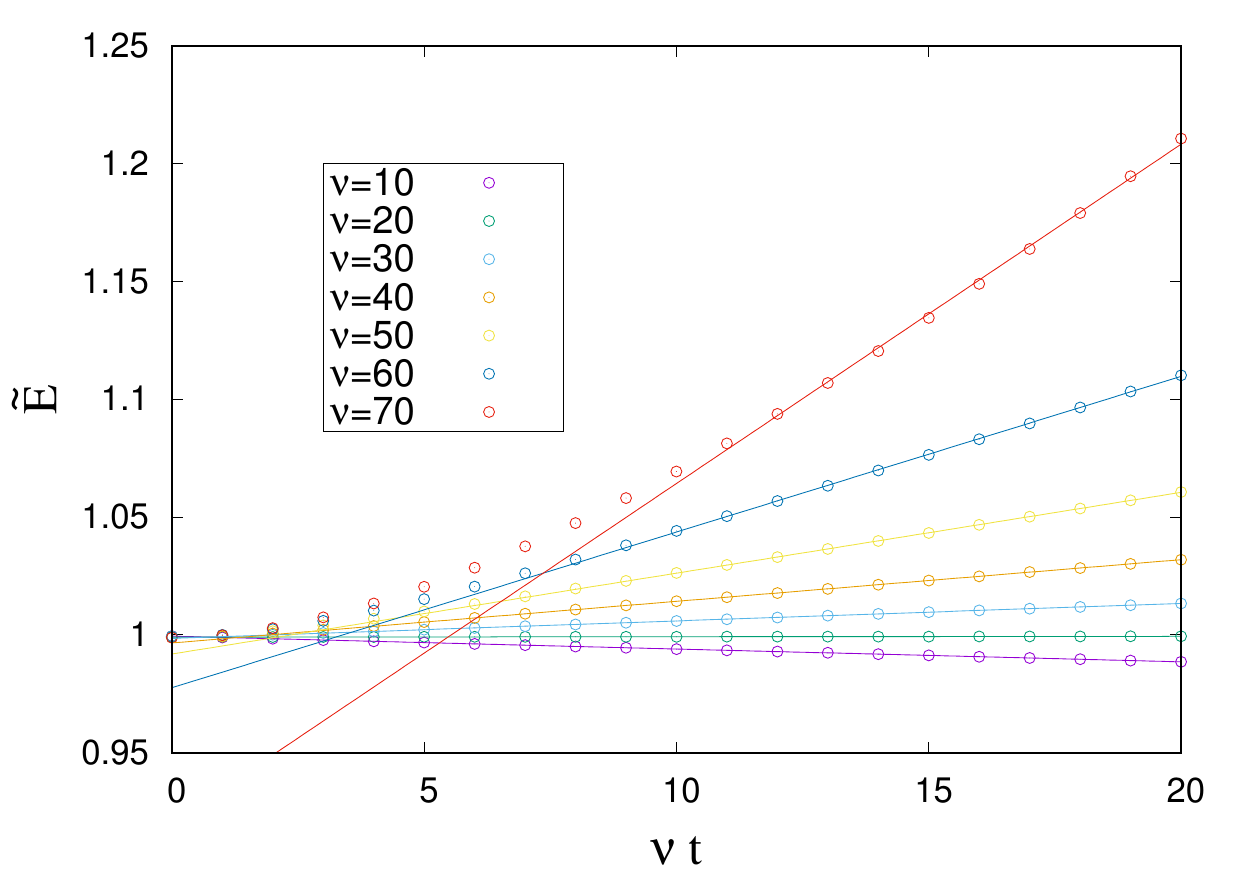}
    \includegraphics[width=0.7\textwidth]{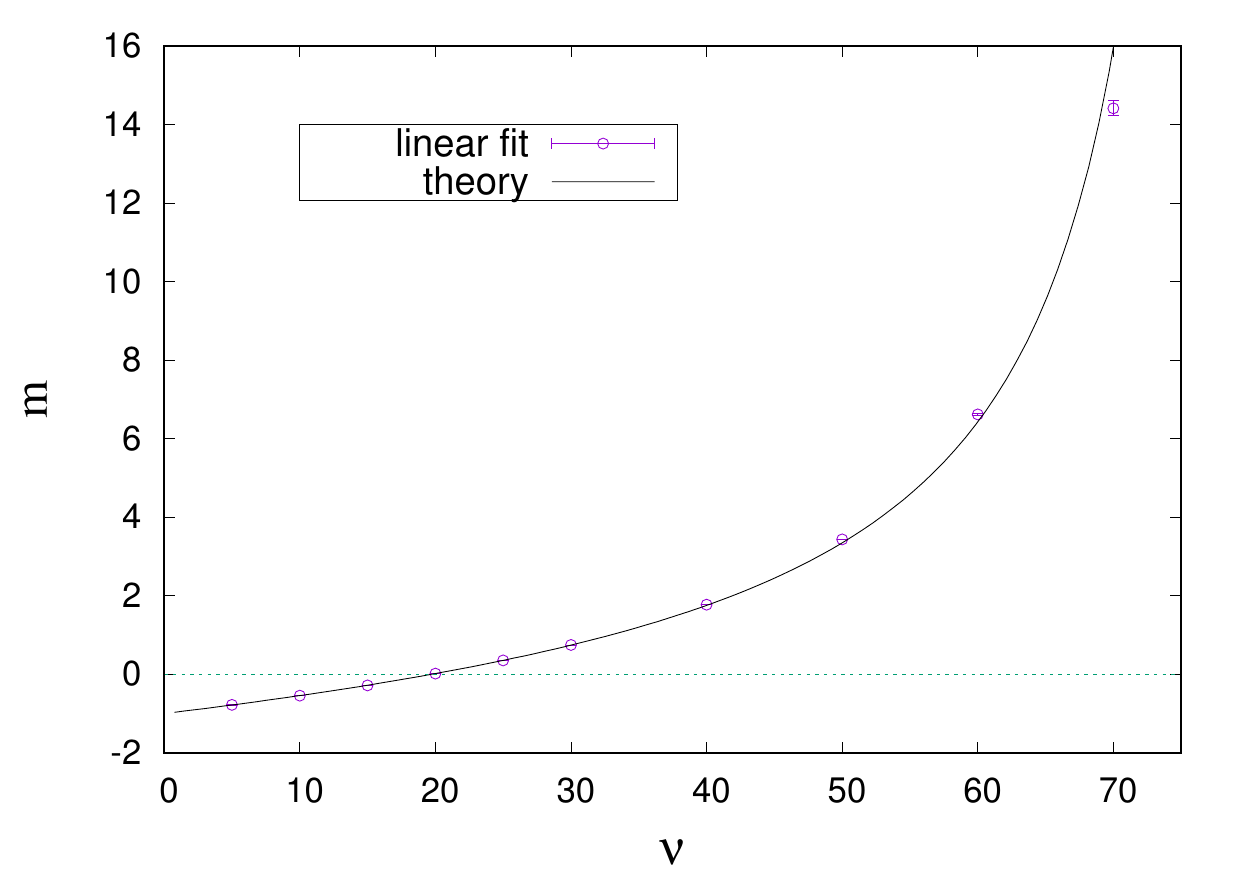}
    \caption{\small Top panel: rescaled energy $\tilde{E}=1+\delta\tilde{E}$ as a function of time. We compare the
      numerical values of $1+\delta\tilde{E}$ (circles) and the linear
      fits (lines) in the second part of the trajectory, for several
      values of $\nu$ (see legend).  Bottom panel: Plot of the slope
      $m=L\,d\tilde{E}/d(\nu t)$ as a function of $\nu$. We
      compare the fitted slopes in the top panel (circles) and their
      theoretical values, as given by $\psi_{\HCS}$
      in~\eqref{eq:total-energy-sc-sol} (blue line). The transition at
      $\nu_{\psi} =\nu_{c}/4= 2 \pi^2$ is marked by the horizontal
      dashed line. We have used a system of size $L=1000$.
}
    
\label{f:haffviolation}
\end{figure}

We compare the theoretical result~\eqref{eq:total-energy-sc-sol} to
Monte Carlo simulations of the model in figure
\ref{f:haffviolation}. This is carried out by fitting
$\delta{\tilde{E}}(t)$ with a straight line in the second part of the
trajectory, that is, for times long enough so as to the velocity
correlations being described by their asymptotic expression
\eqref{eq:reg-HCS} but small as compared to the system size. The
agreement is excellent.

There is a critical dissipation value $\nu_{\psi}$ such that
$\psi_{\HCS}$ vanishes, i.e.
\begin{equation}\label{eq:nu-psi}
\nu_{\psi}=\nu_{c}/4=2\pi^{2}, \qquad \psi_{\HCS}(\nu_{\psi})=0,
\end{equation}
and the finite-size correction in~\eqref{eq:total-energy-sc-sol}
changes sign. Therefore, at this point we find a change in the
time-derivative of $\delta\tilde{E}(t)$.  For large system sizes, the
energy decays faster (slower) than the Haff law for $\nu<\nu_{\psi}$
($\nu>\nu_{\psi}$) because $\psi_{\HCS}<0$ ($\psi_{\HCS}>0$). In the
bottom panel of the figure, we compare the numerical slopes of the 
rescaled temperature with the theoretical prediction $\psi_{\HCS}$ 
as a function of $\nu$. Note that $\psi_{\HCS}$ diverges as $\nu\to\nu_{c}$, 
which is a signature of the shear instability of the HCS.

\section{Beyond Molecular Chaos: Velocity Correlations in the HCS}
\label{sec:beyond_mol_chaos}

In this section, we analyse the effect on the free cooling of the
system introduced by the velocity correlations. The average equation
for the granular temperature (or the energy) in the HCS is closed only
when the correlation $\langle v_{l}v_{l+1}\rangle$ is neglected, since
it is expected to be of the order of $L^{-1}$. In other words, the
evolution equation for the temperature is closed in the Molecular
Chaos approximation. Interestingly, for the case of Maxwell
  molecules we are considering in the paper, we can account for the
effect of the correlations in the cooling of the system in quite a
detailed way.

\subsection{Perturbative Solution for Temperature and Correlations}
\label{sec:beyond_mol_chaos_linear}

We assume that the system is in a spatial-translation-invariant state,
such as the HCS. We define the set of spatial
correlations of the velocity at time $\tau$ as
\begin{equation}\label{eq:C-k}
C_{k}(\tau)=\langle v_{j}(\tau)v_{j+k}(\tau)\rangle.
\end{equation} 
Here, $k$ represents the distance between the involved sites in the
correlation. Note that the average temperature at any site $j$ is given by $C_{0}$,
\begin{equation}
  \label{eq:C0-energy}
  T(\tau)\equiv C_{0}(\tau)=\langle v_{j}^{2}(\tau)\rangle.
\end{equation}
The evolution equation of these correlations is readily
obtained from the master equation,
\begin{subequations}\label{hier}
\begin{eqnarray}
\omega^{-1}\tauder{C_{0}} & = & (\alpha^2-1)(C_{0}-C_{1}), \label{hier1} \\
\omega^{-1}\tauder{C_{1}} & = & (1-\alpha^2)(C_{0}-C_{1})+(1+\alpha)(C_{2}-C_{1}), \label{hier2} \\
\omega^{-1}\tauder{C_{k}} & = &(1+\alpha)(C_{k+1}+C_{k-1}-2C_{k}), \quad 2\leq k\leq (L-1)/2, \label{hier3} \\
\hspace{7mm} C_{\frac{L+1}{2}}&= & C_{\frac{L-1}{2}}, \quad \forall\tau. \label{hier4}
\end{eqnarray}
\end{subequations}
In the above equations, we have omitted the
$\tau$-dependence of the correlations to keep our notation simple. We
have written them for odd $L$, because the ``upper'' boundary
condition (for the maximum value of $k$) is simpler to write. Of
course, this choice is irrelevant in the large system size limit.

As a consequence of momentum conservation, in the center of mass frame
we have the ``sum rule''
\begin{equation}\label{eq:sum-rule}
C_{0}(\tau)+2 \sum_{k=1}^{\frac{L-1}{2}} C_{k}(\tau)=0, \quad \forall\tau.
\end{equation}
For a conservative ($\alpha=1$) system in equilibrium, the
correlations $C_{k}$ do not depend on the distance between sites $k$
and they  are of the order $O(L^{-1})$: $C_{k}^{\eq}=-T(L-1)^{-1}$,
$\forall k> 0$. In a non-equilibrium state, we may have a non-trivial
space structure in the correlations, but we still assume them to be of
the order of $L^{-1}$. Then, we define the rescaled correlations $D_{k}(\tau)$ as
\begin{equation}
  \label{eq:D-def}
  D_{k}(\tau)= L C_{k}(\tau), 
\end{equation}
which we assume to be of the order of unity in the infinite size limit
as $L\to\infty$.

Let us write~(\ref{hier}) in the large system size
limit, in which we expect $D_{k}(\tau)$ to be a smooth function of space, in the sense that
$D_{k+1}(\tau)-D_{k}(\tau)=O(L^{-1})$. Then, the typical
hydrodynamic length and time scales \cite{first} are introduced as
\begin{equation}
  \label{eq:hydro-scales}
  x=\frac{k-1}{L}, \quad t=\frac{\omega\tau}{L^{2}}.
\end{equation}
Keeping solely terms up to
$O(L^{-1})$, we arrive at
\begin{subequations}\label{contcorr}
\begin{eqnarray}
&&\frac{dT(t)}{dt}=-\nu \left[ T(t)-L^{-1}
   \psi(t)\right], \label{contcorr1}\\
&& \nu
   T(t)+4\xder{D(x,t)}|_{x=0}=L^{-1}\left(\frac{d\psi(t)}{dt}+
\nu \psi(t)\right), \label{contcorr2} \\
&&\tder{D(x,t)}=2\,\partial_{xx} D(x,t), \label{contcorr3}
  \\
&&\xder{D(x,t)}|_{x=1/2}=\frac{1}{2}L^{-1}\frac{d\chi(t)}{dt} \label{contcorr4}
\end{eqnarray}
\end{subequations}
in which we have introduced the notations
\begin{equation}\label{eq:psi}
\psi(t)=\lim_{x \rightarrow 0}D(x,t), \quad \chi(t)=\lim_{x \rightarrow\frac{1}{2}}D(x,t).
\end{equation}
These equations are exact up to times such that $t\ll L^{2}$, since
the lowest order terms that have been neglected are of the order of
$L^{-2}$, for instance the fourth-spatial-derivative term in the
diffusion equation (\ref{contcorr2}) for the correlations. In
(\ref{contcorr1}), we have a $L^{-1}$ correction to the cooling rate,
brought about by the nearest-neighbour velocity
correlation.

Of course, these equations are compatible with the sum rule
\eqref{eq:sum-rule}. When we retain only terms up to and including
$O(L^{-1})$, we have 
\begin{equation}
  \label{eq:sum-rule-continuum}
T(t)+2\int_{0}^{1} dx \, D(x,t)+L^{-1}\left[\psi(t)-2\chi(t)\right]=O(L^{-2}),
\end{equation}
as shown in~\ref{app-b}. The lhs of \eqref{eq:sum-rule-continuum} is a constant of motion, as can 
be readily shown by using the evolution equations \eqref{contcorr}. 
 
In order to solve the above system, it is useful to define the scaled
(tilde) fields with their corresponding power of
$T_{\HCS}(t)$. Namely, we define
\begin{equation}
  \label{eq:scaled-corr-apm}
  \tilde{T}(t)=\frac{T(t)}{T_{\HCS}(t)}, \quad
  \widetilde{D}(x,t)=\frac{D(x,t)}{T_{\HCS}(t)}.
\end{equation}
These rescaled fields obey the equations
\begin{subequations}\label{eq:scaled-eqs-apm}
\begin{eqnarray} 
  &&\frac{d\tilde{T}(t)}{dt}=\nu L^{-1}
     \tilde{\psi}(t),   \label{eq:scaled-eqs-apm-1}\\
&& \nu\tilde{T}(t)+4\xder{\widetilde{D}(x,t)}|_{x=0}=L^{-1}\frac{d\widetilde{\psi}(t)}{dt}, \label{eq:scaled-eqs-apm-2} \\
  &&\tder{\widetilde{D}(x,t)}=\nu \widetilde{D}(x,t)+2\,\partial_{xx}
     \widetilde{D}(x,t), \label{eq:scaled-eqs-apm-3}
  \\
  && \xder{\widetilde{D}(x,t)}|_{x=1/2}=\frac{1}{2}L^{-1}\left(\frac{d\widetilde{\chi}(t)}{dt}-\nu \widetilde{\chi}(t)\right). \label{eq:scaled-eqs-apm-4}
\end{eqnarray}
\end{subequations}
The system above is linear in $(\tilde{T},\widetilde{D})$, so it is possible to seek the
exact solution thereof. In fact, we find the exact solution of the
discrete hierarchy \eqref{hier} in section \ref{sec:HCS-exact}. Here,
we are interested in finding the corrections to the cooling rate
introduced by the velocity correlations, so we look for a solution of 
\eqref{eq:scaled-eqs-apm} by means of a perturbative approach. 
This can be performed by expanding all functions of time in powers of $L^{-1}$,
\begin{subequations}\label{eq:1/L-expansion}
\begin{equation} \tilde{T}(t)=\tilde{T}_0(t)+L^{-1}\tilde{T}_1(t)+O(L^{-2}),
\end{equation}
\begin{equation} \widetilde{D}(x,t)=\widetilde{D}_0(x,t)+L^{-1}\widetilde{D}_1(x,t)+O(L^{-2}),
\end{equation}
\end{subequations}
with analogous expansions for $\widetilde{\psi}(t)$ and $\widetilde{\chi}(t)$.\\
To the lowest order, we have
\begin{subequations}\label{eq:0-order}
\begin{eqnarray} 
  &&\frac{d}{dt}\tilde{T}_0 =0,   \label{eq:0-order-1}\\
&& \nu\tilde{T}_0+4\xder{\widetilde{D}_0}|_{x=0}=0, \label{eq:0-order-2} \\
  &&\tder{\widetilde{D}_0}=\nu \widetilde{D}_0+2\,\partial_{xx}
     \widetilde{D}_0, \label{eq:0-order-3}
  \\
  && \xder{\widetilde{D}_0}|_{x=1/2}=0. \label{eq:0-order-4}
\end{eqnarray}
\end{subequations}
From~\eqref{eq:0-order-1}, we have that $\tilde{T}_0=1$ is a constant. 
Moreover, in the limit $t\!\gg\!\! 1$, the scaled correlations tend to a
stationary value, which is given by
\begin{equation}\label{eq:stcorr1}
  \widetilde{D}_0(x)= - A \cos
\left[\pi\sqrt{\frac{\nu}{\nu_{c}}} (1 - 2x) \right], \quad A
=\pi\sqrt{\frac{\nu}{\nu_{c}}}\csc\left(\pi\sqrt{\frac{\nu}{\nu_{c}}}\right).
\end{equation}
Looking for the first order corrections, for our purposes we only need to write the evolution 
equation for $\tilde{T_1}(t)$,
\begin{equation}
\frac{d}{dt}\tilde{T}_1(t) = \nu \widetilde{\psi}_0 (t) 
\label{eq:T1}
\end{equation}
hence when the correlations reached the stationary profile~\eqref{eq:stcorr1} 
we have that
\begin{equation}
\frac{d}{dt}\tilde{T}_1(t) = \nu \psi_{\HCS}
\label{eq:stT1}
\end{equation}
where $\psi_{\HCS}$ is the same quantity that we defined in \eqref{eq:reg-psi-HCS} 
within the mesoscopic fluctuation theory framework, which we rewrite as
\begin{equation}
  \label{eq:D0-psi}
   \quad 
\psi_{\HCS}=-\pi \sqrt{\frac{\nu}{\nu_{c}}} \cot \left(\pi\sqrt{\frac{\nu}{\nu_{c}}}  \right)  .
\end{equation}
Therefore, for $t\!\gg\!\! 1$ the rescaled temperature is linearly diverging as
\begin{equation}
\tilde{T} (t) \sim  1 + \frac{\nu \psi_{\HCS}}{L} t + O(L^{-2})
\label{eq:Ttilde}
\end{equation}
neglecting the transient terms for $\tilde{T}_1$. This result is equivalent to the one 
in~\eqref{eq:total-energy-sc-sol} (known that $E(t)=T(t)$ in the homogeneous case) 
and has been compared with simulations in figure \ref{f:haffviolation}.

\begin{figure}[!ht]
\centering
\includegraphics[width=0.7\textwidth]{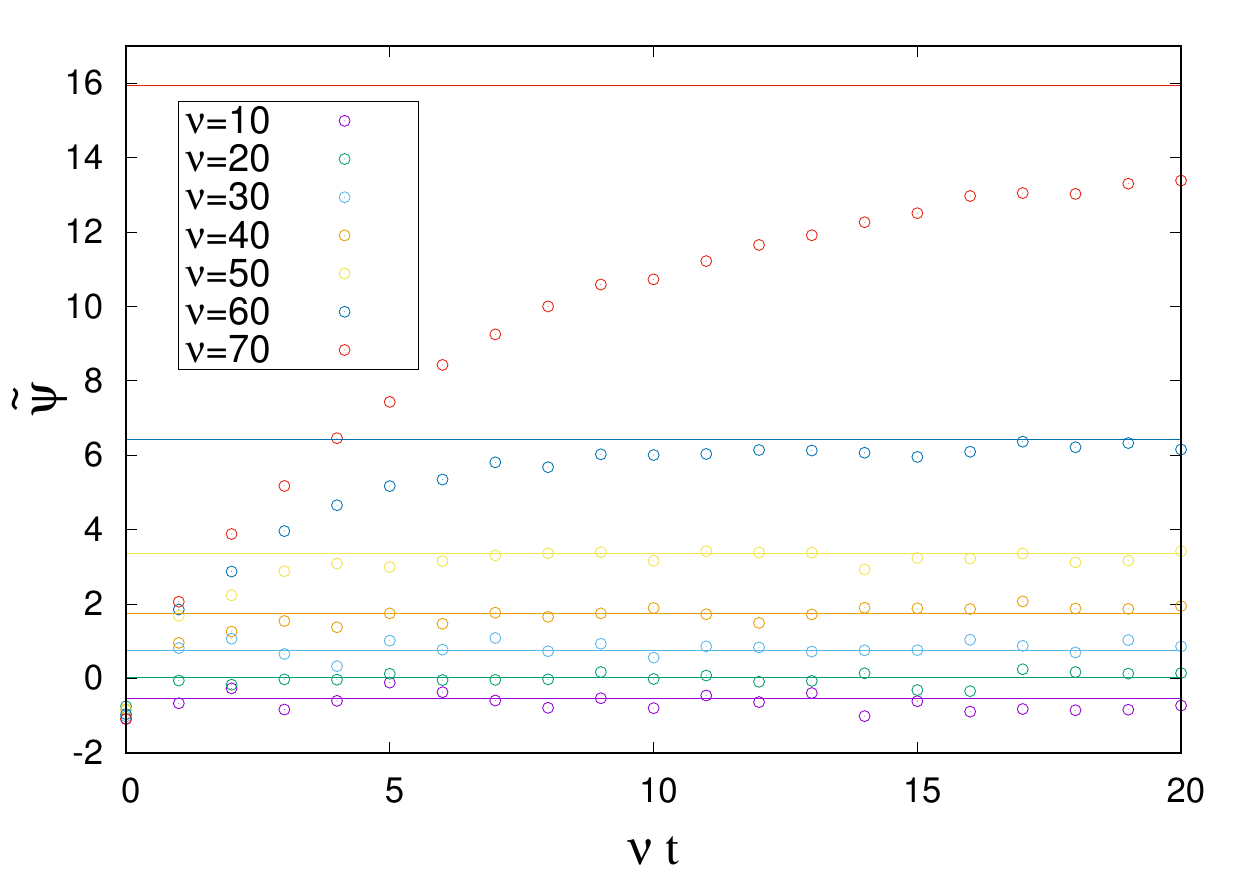}
\includegraphics[width=0.7\textwidth]{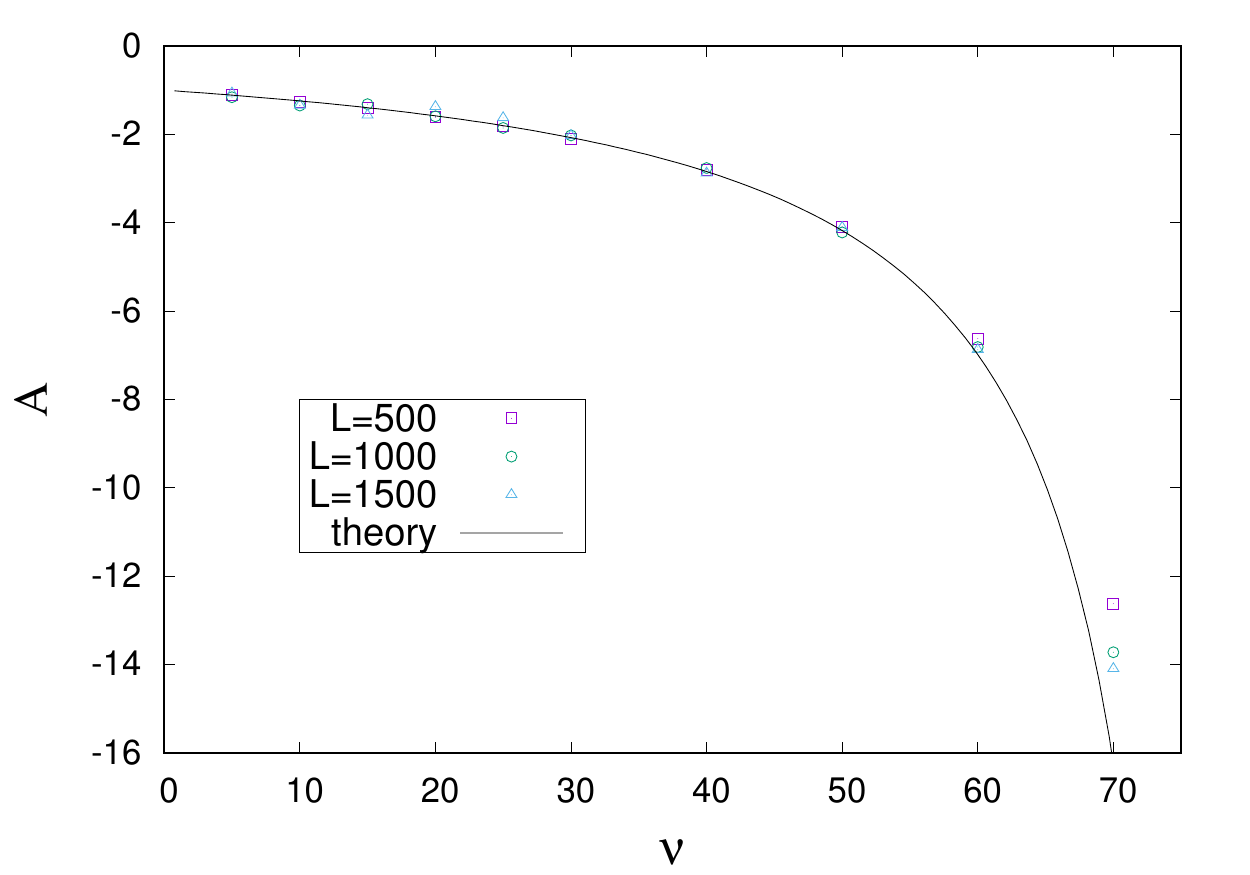}
\caption{\small Top panel: Time evolution of nearest-neighbour
  correlations $\psi(t)$.  We plot their numerical values (circles)
  for several $\nu$ (see legend) and $L=1000$ particles as a function of
  the dimensionless time $\nu t$, and their theoretical stationary
  values, given by~\eqref{eq:D0-psi}.  In the plotted time window, the
  correlations reach their stationary value for all $\nu\leq60$, while
  they do not for $\nu=70$. This discrepancy will be analysed in
  figure~\ref{f:haff-mult-scale}. Bottom panel: Correlation amplitude
  $A$, defined in~\eqref{eq:stcorr1}, as a function of $\nu$. We plot
  both its numerical value, computed in simulations for
  $L=250,500,1000$ (symbols), and its theoretical expectation (black
  line). A very good agreement is found for all
  $\nu<70$. 
}
\label{f:amplitudes}
\end{figure}

We have also checked the theoretical prediction~\eqref{eq:stcorr1} for the velocity 
correlations in the HCS in figure~\ref{f:amplitudes}. Thus, we plot the simulation value 
of the amplitude $A$ as a function of $\nu$, and compare it with~\eqref{eq:stcorr1}. 
Trajectories start from a homogeneous mesoscopic velocity profile with zero average, 
$u(x,0) \equiv 0$. Once more, a very good agreement is found.

We already commented that the result in \eqref{eq:Ttilde} is valid
only for $\psi_{\HCS} \nu t /L \ll 1$, while in this section we used
the stationary value of the correlations supposing $t \gg 1$.
Depending on the value of $\nu$ and $L$, these conditions on time
  may be either consistent or inconsistent. In fact, numeric data in
figure~\ref{f:amplitudes} show an excellent agreement with the theoretical
prediction in~\eqref{eq:Ttilde} for $\nu<60$, while for higher
dissipation the nearest-neighbour correlations do not seem to have
reached their stationary value. Therefore, longer trajectories
should be observed and this leads to the divergence of the first order
perturbation $O(t/L)$.

\subsection{Temperature and Correlations Evolution: Multiple-Scale Analysis}
\label{sec:beyond_mol_chaos_msa}

In order to build up a theory which give a consistent picture
  for long times, we introduce a multiple-scale perturbative solution
of \eqref{eq:scaled-eqs-apm}. Equation
\eqref{eq:scaled-eqs-apm-1} suggests the introduction of two distinct
time scales: apart from $t$, we define a slow time scale $\sigma$,
\begin{equation} 
\label{eq:slow-scale}
s=t, \; \sigma=L^{-1}t, \quad
\tder=\partial_{s}+L^{-1}\partial_{\sigma}.
\end{equation}
Our notation makes it possible to distinguish between $\partial_{t}$
(with constant $x$) and $\partial_{s}$ (with constant $x$ and
$\sigma$). All functions of time are expanded in powers of $L^{-1}$ as before and considered 
to depend on both time scales $(s,\sigma)$. So, to the lowest order we have
\begin{subequations}\label{eq:0-order-msa}
\begin{eqnarray} 
  &&\sder{\tilde{T}_0}(s,\sigma)=0,   \label{eq:0-order-1-msa}\\
&& \nu\tilde{T}_0+4\xder{\widetilde{D}_0}|_{x=0}=0, \label{eq:0-order-2-msa} \\
  &&\sder{\widetilde{D}_0}=\nu \widetilde{D}_0+2\,\partial_{xx}
     \widetilde{D}_0, \label{eq:0-order-3-msa}
  \\
  && \xder{\widetilde{D}_0}|_{x=1/2}=0. \label{eq:0-order-4-msa}
\end{eqnarray}
\end{subequations}
which has the same form of~\eqref{eq:0-order} but now $\tilde{T}_0$
depends also on the slow time scale $\sigma$; more precisely,
from~\eqref{eq:0-order-1-msa} we have that it depends only on
$\sigma$,
$\tilde{T}_0(\cancel{s},\sigma)=\tilde{T}_0(\sigma)$. Note
  that $\tilde{T}_{0}(\sigma)$ remains undetermined at the lowest
order. Also, \eqref{eq:0-order-msa} leads now to a
pseudo-stationary solution for $\widetilde{D}_0(x,\cancel{s},\sigma)$
for long time scales $s \gg 1$ but finite $\sigma$, namely
\begin{equation}\label{eq:stcorr1-msa}
  \widetilde{D}_0(x,\sigma)= - \tilde{T}_0(\sigma) A \cos
\left[\pi\sqrt{\frac{\nu}{\nu_{c}}} (1 - 2x) \right] ,
\end{equation}
which differs from~\eqref{eq:stcorr1} because of the $\sigma$ dependence 
of $\tilde{T}_0(\sigma)$. As is usual in multiple-scale analysis, the 
latter can be obtained by writing down the equations for the first order 
corrections. In fact, for the purposes of the present paper, it suffices 
to write the evolution equation for  $T_1(s,\sigma)$,
\begin{equation}
  \label{eq:eps-1-msa}
\sder{\tilde{T}}_1 + \sigdertot{\tilde{T}_0}=\nu \widetilde{\psi}_0, \qquad
 \widetilde{\psi}_0 = \tilde{T}_0\psi_{\HCS}.
\end{equation}
Since the rescaled energy should not contain linear terms in time (see
section \ref{sec:HCS-exact} for a rigorous proof), the first lhs term
of~\eqref{eq:eps-1-msa} must vanish, and
\begin{equation}
  \label{eq:3}
  \nu\psi_{\HCS} \tilde{T}_0(\sigma)-\sigdertot{\tilde{T}_0}=0 \;
  \Rightarrow \; \tilde{T}_0(\sigma)=e^{\nu\psi_{\HCS}\sigma},
\end{equation}
where we have taken into account that $\tilde{T}_0(t=0)=1$. Going back to 
the unscaled variables, what we have shown is that
\begin{equation}
  \label{eq:haff-renorm}
 T(t)=T(0) \exp\left[-\nu_{\HCS}^{r}t\right]+O(L^{-1}), \qquad  \nu_{\HCS}^{r}=\nu\left(1-L^{-1}\psi_{\HCS}\right) .
\end{equation}
Equation~(\ref{eq:haff-renorm}) tells us that the cooling rate in Haff's law has a finite 
size correction. Of course, if we consider that $\sigma=t/L\ll 1$ and retain only the 
linear terms in $L^{-1}$, we reobtain the results in section \ref{sec:meso-fluc-th} and 
in~\eqref{eq:Ttilde}. 
\begin{figure}[!ht]
\centering
\includegraphics[width=0.7\textwidth]{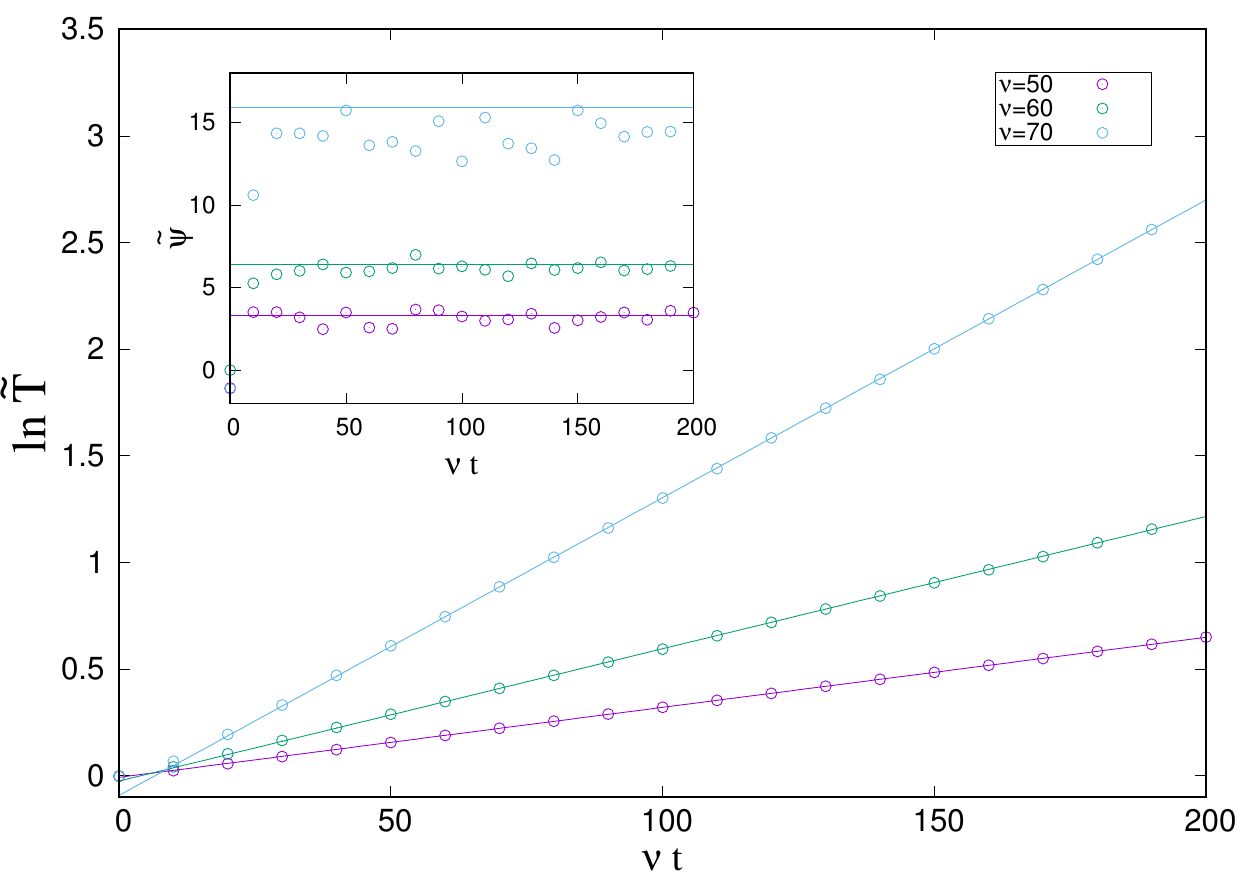}
\includegraphics[width=0.7\textwidth]{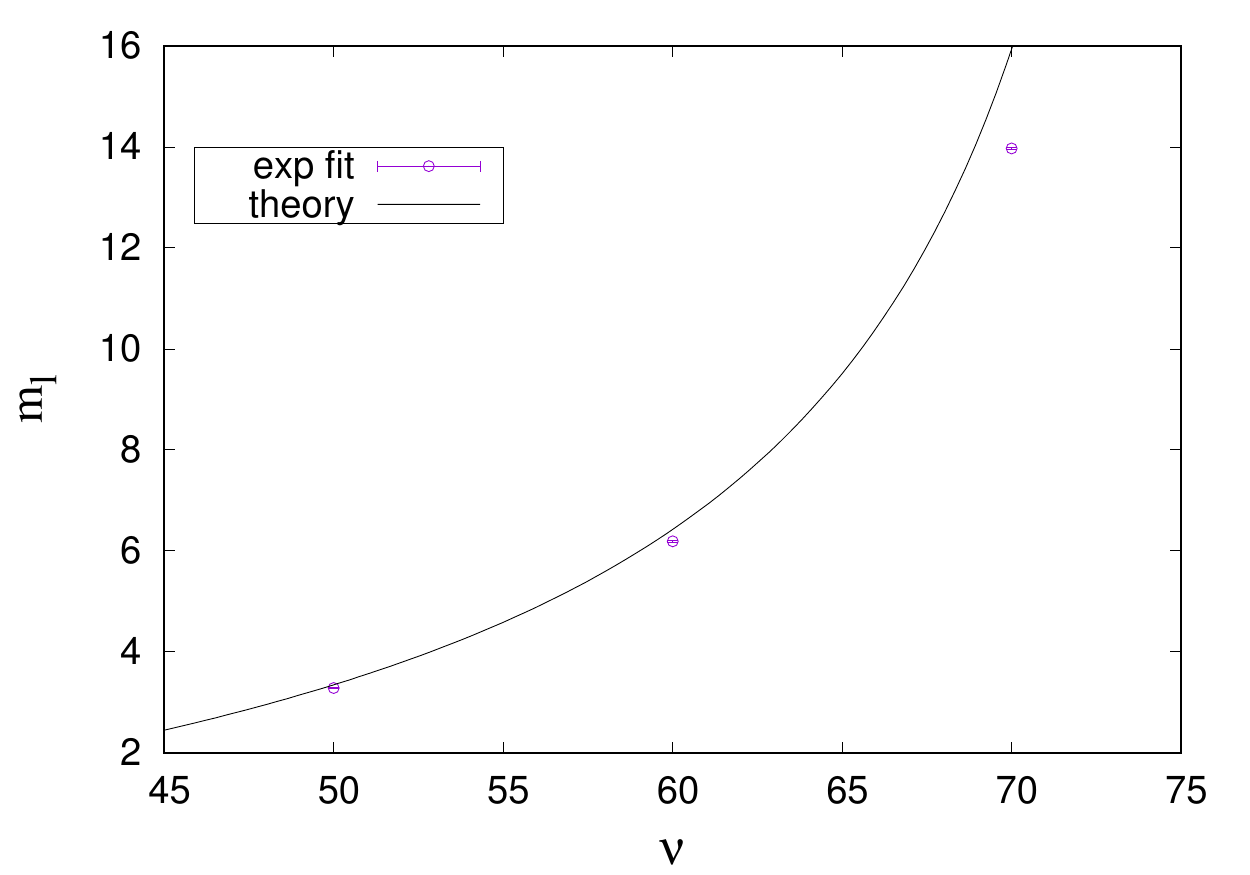}
\caption{\small Top panel: Log-linear plot of the rescaled
  temperature.  The simulation values are plotted for $\nu=50,60,70$
  (circles), and also the fits made upon the second part of the long
  trajectories (lines). The time evolution is clearly exponential as
  predicted from~\eqref{eq:3}. Inset: Time
  evolution of the nearest-neighbour correlations $\psi(t)$ for long
  trajectories. We have plotted the simulation curves (circles) and
  their corresponding theoretical stationary values (lines).
  Bottom panel: Slope $m_l$
  of the time evolution of $\ln \tilde{T}$. The fitting values from
  the top panel (squares) are plotted together with the
  theoretical prediction~\eqref{eq:3} (black line). All the
  trajectories have been done with $L=1000$ particles up to a
  maximum time $\nu t = 200$. 
}
\label{f:haff-mult-scale}
\end{figure}

We check the renormalisation of Haff's law predicted by
\eqref{eq:haff-renorm} in figure~\ref{f:haff-mult-scale}: simulations
made over long times $\nu t \psi_{\HCS} \sim L$ show that the rescaled
temperature has an exponential behaviour, as predicted from the
multiple-scale analysis. The exponential slope has been fitted and
numerical results are in good agreement with the theoretical
prediction~\eqref{eq:3}. Nearest-neighbour correlations have been
studied as before: figure~\ref{f:haff-mult-scale} shows that for
$\nu=50,60$ they converge to their expected value after a very short
transient, whereas for $\nu=70$ they also converge but to a stationary
value smaller than the expected one. This effect is probably given by
next order corrections which are becoming relevant when approaching
the critical dissipation $\nu_c$, where we know that $\psi_{\HCS}$ is
divergent.

\section{Exact solution of the HCS in a finite system}\label{sec:HCS-exact}

The hierarchy~\eqref{hier} can be exactly solved by reducing it to the
eigenvalue problem of a certain matrix. As before, we carry out
this approach to the problem for odd $L$; a choice that is
irrelevant in the large system size limit $L\gg 1$.  The problem for
an even number particles may be solved by following an utterly similar
strategy, but the boundary conditions are a little more involved to
write. We do not present here these calculations because they do not
provide any additional physical insight.

First, it is useful to introduce a change of variables in order to make
the matrix symmetric. Specifically, we define
\begin{equation}
c_0=C_0 , \qquad c_k=\sqrt{2}\,C_k , \quad 1 \leq k \leq (L-1)/2.
\end{equation}
Second, we rewrite the  hierarchy~\eqref{hier} as
\begin{subequations}\label{hierv2}
\begin{eqnarray}
\hspace{3mm} \omega^{-1}(1+\alpha)^{-1}\partial_{\tau} c_{0} & = & -(1-\alpha)c_{0}+ \frac{1-\alpha}{\sqrt{2}}c_{1}, \label{hierv21} \\
\hspace{3mm} \omega^{-1}(1+\alpha)^{-1}\partial_{\tau}c_{1} & = & \frac{1-\alpha}{\sqrt{2}} c_{0}-\frac{3-\alpha}{2}c_1+c_2, \label{hierv22} \\
\hspace{3mm} \omega^{-1}(1+\alpha)^{-1}\partial_{\tau} c_{k} & = & c_{k-1}-2c_{k}+c_{k+1}, \quad 2\leq k\leq (L-3)/2, \label{hierv23} \\
\omega^{-1}(1+\alpha)^{-1}\partial_{\tau} c_{\frac{L-1}{2}} & = & c_{\frac{L-3}{2}} - c_{\frac{L-1}{2}}, \label{hierv24}
\end{eqnarray}
\end{subequations}
in which we have extracted the common factor $(1+\alpha)$ on the rhs
of \eqref{hier} and made use of \eqref{hier4} to write \eqref{hierv24}
for $c_{\frac{L-1}{2}}$. 

Now, we can solve the system above by a standard eigenvector method,
that is, we seek solutions of the form
\begin{equation}
c_k=e^{\lambda (1+\alpha)\omega\tau} \phi_k.
\end{equation} 
We denote the eigenvalues by $\lambda$ and its corresponding eigenvector by
$\phi$, $\phi_{k}$ is thus the $k$-th component thereof. In this way, we reach the system
\begin{subequations}\label{eigen}
\begin{eqnarray}
\hspace{3mm} \lambda \phi_{0} & = & -(1-\alpha) \phi_{0}+ \frac{(1-\alpha)}{\sqrt{2}} \phi_{1}, \label{eigen1} \\
\hspace{3mm} \lambda \phi_{1} & = & \frac{(1-\alpha)}{\sqrt{2}} \phi_{0}-\frac{(3-\alpha)}{2} \phi_1+ \phi_2, \label{eigen2} \\
\hspace{3mm} \lambda \phi_{k} & = & \phi_{k-1}-2 \phi_{k}+ \phi_{k+1}, \quad 2\leq k\leq (L-3)/2, \label{eigen3} \\
\lambda \phi_{\frac{L-1}{2}} & = & \phi_{\frac{L-3}{2}} - \phi_{\frac{L-1}{2}}. \label{eigen4}
\end{eqnarray}
\end{subequations}

Equations \eqref{eigen} are a system of second-order difference
equations for $\phi_k$ with contant coefficients, in
which~\eqref{eigen3} is the general equation and \eqref{eigen2} and
\eqref{eigen4} are their boundary conditions. On top of that,
\eqref{eigen1} acts as an extra condition that ensures momentum
conservation, as shown below (see also section
\ref{sec:beyond_mol_chaos}). The general solution of~\eqref{eigen3} is
of the form $\phi_{k>0}=r^k$ \cite{Be99}, which substituted into
\eqref{eigen3} has two solutions $(r_1,r_2)$ that verify
\begin{equation}\label{eigenr}
 r_1 r_2  =  1, \qquad
r_1 + r_2  =  2 + \lambda. 
\end{equation}
We introduce a new variable $q\in[0,\pi]$ such that $r_1=e^{iq}$ and
$r_2=e^{-iq}$, as suggested by \eqref{eigenr}. Note that
$|r_{1}|=|r_{2}|=1$, if one of the roots were larger than one it would
lead to correlations increasing with $k$, which is physically
absurd. Moreover, from a purely mathematical point of view,
restricting ourselves to $|r_{1}|=|r_{2}|=1$ leads to a complete set
of eigenvectors.  From \eqref{eigenr}, we obtain
\begin{equation}
\lambda(q)=2(\cos q - 1),
\end{equation} 
and the corresponding eigenvector is given by
\begin{subequations}\label{eigenphi}
\begin{eqnarray}
\phi_{k>0}(q) & = & A \,e^{ikq} + B \, e^{-ikq}, \\
\hspace{3mm} \phi_{0}(q) & = & \frac{1-\alpha}{\sqrt{2} \left(2\cos q-1-\alpha\right)}  (A\, e^{iq} + B\, e^{-iq}).
\end{eqnarray}
\end{subequations}

\begin{figure}
  \centering
  \includegraphics[width=0.7\textwidth]{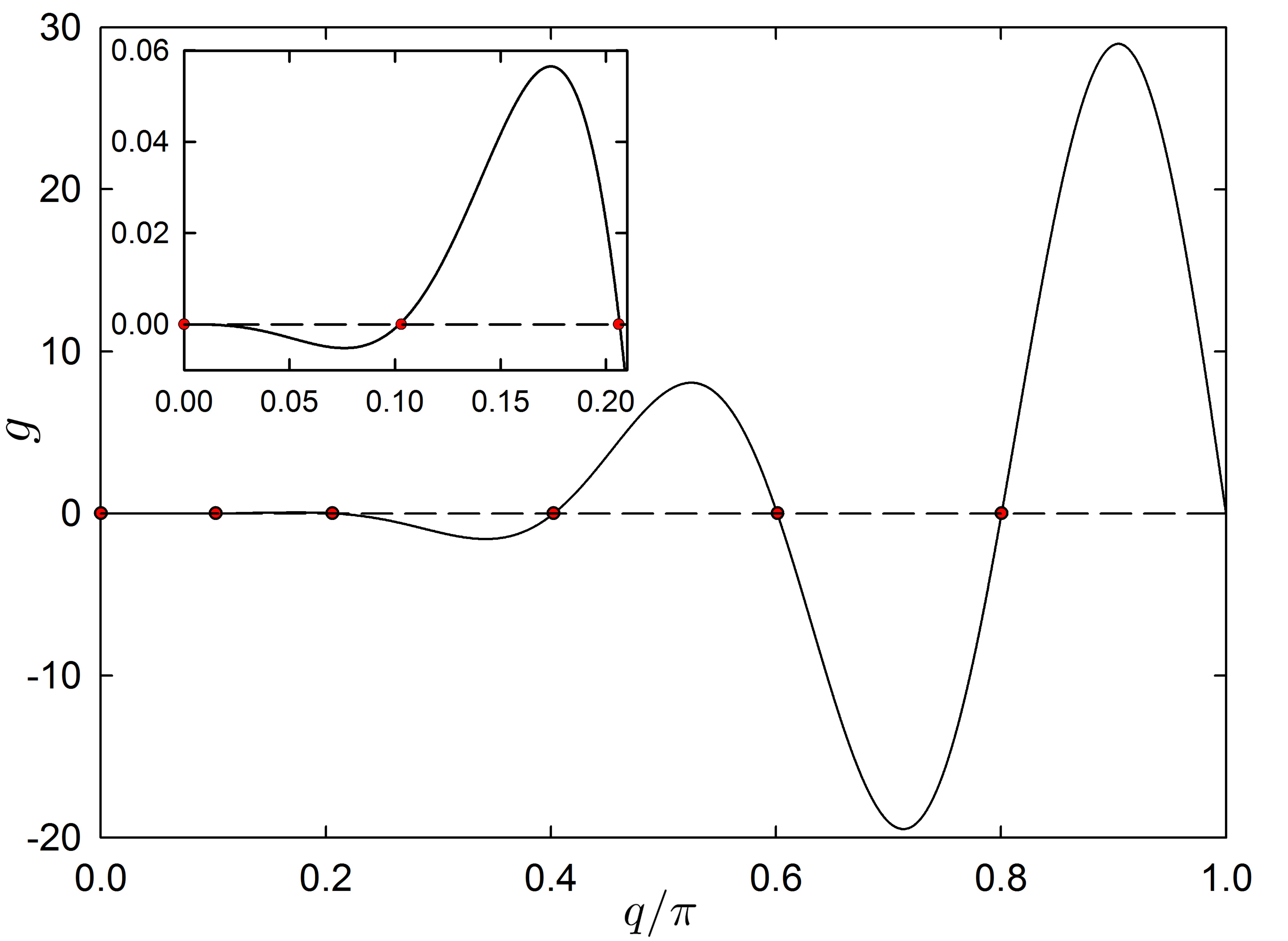}
  \caption{\label{fig:g(q)} Plot of the function $g(q)$ defined
    in~\eqref{eigenfunction} for $L=11$. The zeros $q_{n}$ of this
    function determine the eigenvalues, by taking into account
    \eqref{eq:lambda-qn}. The first zero is always $q_{0}=0$, and
    there are $(L-1)/2$ non-vanishing zeros $q_{i}$,
    $i=1,\ldots,(L-1)/2$.  A zoom of the small $q$ region is shown in
    the inset, in order to make it easier the identification of
    $q_{1}$ and $q_{2}$.}
\end{figure}

The boundary conditions~\eqref{eigen2} and \eqref{eigen4} determine
the constants $A$ and $B$, and also the allowed values of the ``index'' $q$. The determinant of the linear system for
$A$ and $B$ must be zero, which is equivalent to impose that $q$ must
be a zero of the function
\begin{eqnarray}
\label{eigenfunction}
g(q)= & &2 \sin \left( \frac{L+3}{2} q \right) -(5+3\alpha) \sin \left( \frac{L+1}{2} q \right) +(5+7\alpha) \sin \left( \frac{L-1}{2} q \right) \nonumber\\
& &-(3+5\alpha) \sin \left( \frac{L-3}{2} q \right) +(1+\alpha) \sin \left( \frac{L-5}{2} q \right).
\end{eqnarray} 
This function has $(L+1)/2$ different zeros in the half-open interval
$[0,\pi)$, which we denote by $q_{n}$: $q_{0}=0$, $q_n$ is the $n$-th
non-vanishing zero of $g(q)$, $n=1,\ldots, (L-1)/2$. Therefore, we
have found $(L+1)/2$ eigenvalues
\begin{equation}
  \label{eq:lambda-qn}
  \lambda_{n}=2(\cos q_{n}-1),
\end{equation}
the corresponding eigenvectors of which give a complete set for the
problem at hand. In figure~\ref{fig:g(q)} we plot the function $g(q)$
for $L=11$, which has six zeros in the interval $[0,\pi)$.

The eigenvector $\phi^{(n)}$ corresponding to the eigenvalue
$\lambda_n$ can be thus written up to a normalisation constant
$\mathcal{N}_{n}$,
\begin{subequations}\label{eigenvector}
\begin{eqnarray}
\phi^{(n)}_{k>0} & = & \mathcal{N}_n \cos\left[ \left( \frac{L}{2} - k \right) q_n \right], \\
\phi^{(n)}_{0} & = & \frac{(1-\alpha) \mathcal{N}_n}{\sqrt{2} \left(2 \cos q_{n} - 1-\alpha\right)}  \cos\left[ \left( \frac{L}{2} - 1 \right)q_n \right].
\end{eqnarray}
\end{subequations}
The above expressions clearly show that there is no eigenvector for
$q=\pi$, since all its components are zero (recall that $L$ is
odd). The constant $\mathcal{N}_{n}$ is chosen to obtain
a orthonormal set of eigenvectors, in the sense that
\begin{equation}
\sum_{k=0}^{\frac{L-1}{2}} \phi^{(n)}_k \phi^{(n')}_k=\delta_{nn'}. 
\end{equation} 
We do not give the explicit expression for $\mathcal{N}_{n}$ because
it is quite involved and is not necessary for our purposes. The eigenvector corresponding to $q_{0}=0$ is particularly simple,
\eqref{eigenvector} implies that
\begin{equation}
  \label{eq:0-eigenvector}
  \phi_{0}^{(0)}=\frac{\mathcal{N}_{0}}{\sqrt{2}}, \quad
  \phi_{k>0}^{(0)}=\mathcal{N}_{0}, \qquad \mathcal{N}_{0}=\sqrt{\frac{2}{L}}.
\end{equation}
Then, the orthogonality relation of $\phi^{(0)}$ and $\phi^{(n)}$
($n\neq 0$) makes it possible to write a ``sum rule'' for the
components of the latter eigenvectors,
\begin{equation}
  \label{eq:sum-rule-eigenv}
  \phi_{0}^{(n)}+\sqrt{2}\sum_{k=1}^{\frac{L-1}{2}} \phi_{k}^{(n)}=0,
  \quad n>0.
\end{equation}
This sum rule is connected with ~\eqref{eq:sum-rule}, which stemmed
from momentum conservation. It also allows us to write
$\phi_{0}^{(n)}$ in a more convenient form for some calculations,
\begin{equation}
  \label{eq:phi0-conv}
\phi_{0}^{(n)}=-\frac{\mathcal{N}_{n}}{\sqrt{2}}\csc\left(\frac{q_{n}}{2}\right)\sin\left(\frac{L-1}{2}q_{n}\right),
\end{equation}
which does not depend explicitly on $\alpha$.

Finally, we have all the ingredients to build the general solution of \eqref{hierv2} as the sum
\begin{equation}
\label{eigensum}
c_k=\sum_{n=1}^{\frac{L-1}{2}} a_n \,e^{\lambda_n(1+\alpha)\omega\tau} \phi_k^{(n)},
\end{equation}
where $a_n$ is given in terms of the initial conditions by
\begin{equation}
a_n = \sum_{k=0}^{\frac{L-1}{2}} \phi_k^{(n)} c_k(0).
\end{equation}
The sum in~\eqref{eigensum} starts from $n=1$ because $a_{0}=0$, since
\begin{equation}
a_0= \mathcal{N}_0 \left[ \frac{c_0(0)}{\sqrt{2}} + \sum_{k=1}^{\frac{L-1}{2}} c_k(0) \right]=\frac{\mathcal{N}_0}{\sqrt{2}} \left[ C_0(0) + 2 \sum_{k=1}^{\frac{L-1}{2}} C_k(0) \right]=0.
\end{equation}
We have made use of momentum conservation, as expressed by the sum rule \eqref{eq:sum-rule}, in the last equality.

\subsection{Eigenvalues for large systems}

Here, we would like to derive an approximate expression for the
eigenvalue spectrum in the large system size limit $L\gg 1$.
Therefore, we consider that the microscopic dynamics is quasi-elastic
by introducing the macroscopic dissipation coefficient $\nu$,
$(1-\alpha^2)L^2=\nu$, as in \eqref{eq:av-d-nu}. The eigenvalues are
given by the zeros of function \eqref{eigenfunction}, and we expand
this function for $q \ll 1$ by introducing the scaling $Q=qL$, with
the result
\begin{equation}
\tan \left( \frac{Q}{2} \right) \left( \frac{\nu}{2} Q^2 L^{-2} - Q^4 L^{-4} \right) + \frac{1}{2} Q^5 L^{-5} =0.
\end{equation}
We are assuming that $Q$ is of the order of unity and have neglected
terms of the order of $L^{-6}$.

In order to obtain an analytical approximation for the eigenvalues, we
propose an expansion of $Q_{n}=q_{n}L$ in powers of $L^{-1}$,
$Q_n = Q^{(0)}_n+Q^{(1)}_n L^{-1}+O(L^{-2})$. To the lowest order, we
obtain
\begin{subequations}\label{Qzeroth}
\begin{eqnarray}
Q^{(0)}_{1} & = & \sqrt{\frac{\nu}{2}}, \\
Q^{(0)}_{n} & = & 2(n-1) \pi, \qquad n=2,\ldots,(L-1)/2.
\end{eqnarray}
\end{subequations}
Moreover, the  finite size corrections are
\begin{subequations}\label{Qfirst}
\begin{eqnarray}
Q^{(1)}_{1} & = & \frac{\nu}{8 \tan \left( \frac{1}{2}\sqrt{\frac{\nu}{2}} \right)}, \\
Q^{(1)}_{n} & = & \frac{16 (n-1)^3 \pi^3}{8 (n-1)^2 \pi^2 - \nu}, \qquad n=2,\ldots,(L-1)/2.
\end{eqnarray}
\end{subequations}
Note that $Q_{1}^{(1)}$ vanishes at $\nu=\nu_{\psi}=2\pi^{2}$ whereas
it diverges at $\nu=\nu_{c}=8\pi^{2}$. The former property is
connected to the change of sign in the finite-size correction to the
cooling rate of the HCS whereas the latter gives rise to the
instability of the HCS, as discussed in
sections~\ref{sec:meso-fluc-th} and \ref{sec:beyond_mol_chaos}.

\begin{figure}
  \centering
\includegraphics[width=0.7\textwidth]{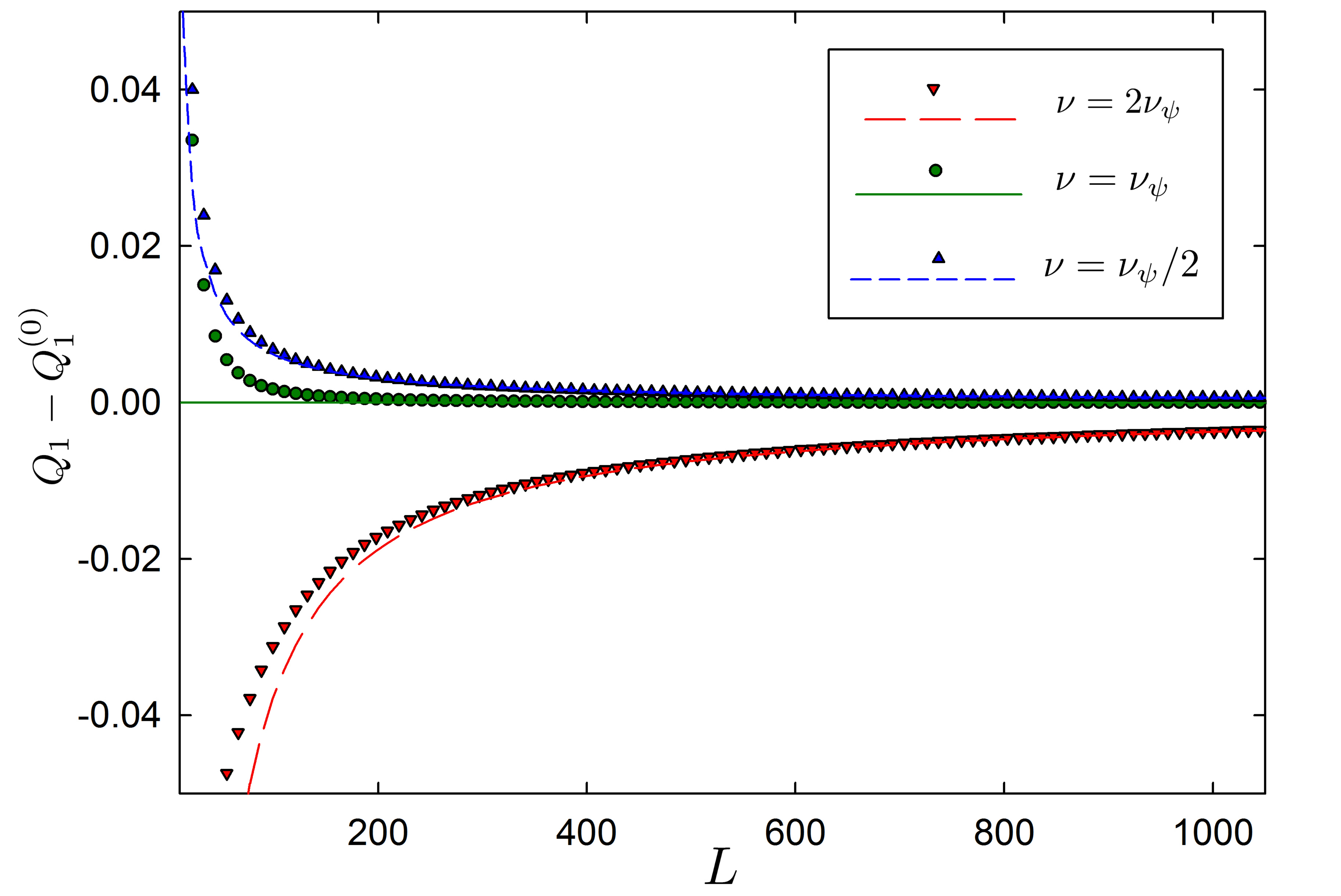}  
  \caption{Plot of the difference $Q_{1}-Q_{1}^{(0)}$ as a function of
  the system size $L$. Three different values of the macroscopic
  dissipation coefficient $\nu$ are considered, namely $\nu=\pi^{2}$,
  $\nu=\nu_{\psi}=2\pi^{2}$ and $\nu=4\pi^{2}$. Two curves are shown
  for each value of $\nu$: the lines correspond to the theoretical
  curve $Q_{1}^{(1)}/L$ and the symbols to the numerical estimate of
  $Q_{1}$ (see the legend for details). It is clearly observed that
  the finite-size correction is especially small for $\nu=\nu_{\psi}$.}
 \label{fig:eigenvalues-expansion} 
\end{figure}

In figure \ref{fig:eigenvalues-expansion}, we check the above
expansion for the zeros of the function $g(q)$. Specifically, we do so
for the first zero $q_{1}$: the numerical estimation of $q_{1}$ is
compared with the expansion in \eqref{Qzeroth} and \eqref{Qfirst} by
plotting $Q_{1}-Q_{1}^{(0)}$ as a function of the system size $L$.  It
is observed that this difference tends to zero as the system size
increases, for all the considered values of $\nu$. The finite size
correction is especially small for $\nu=\nu_{\psi}=2\pi^{2}$, for
which the theoretical correction $Q_{1}^{(1)}$ vanishes. Therefore,
finite size corrections are as small as possible for this case, which
makes it particularly adequate to investigate the hydrodynamic
description, as done in \cite{noi,first}.

We want to emphasise that the instability of the HCS is reobtained
here as a crossing between the first two non-zero eigenvalues: for
$\nu=\nu_c=8\pi^{2}$, we have that $Q_1^{(0)}=Q_2^{(0)}$. On the one
hand, for $\nu<\nu_{c}$, the largest nonvanishing eigenvalue is $\lambda_{1}$
($\lambda_{1}<0$) and dominates the long-time dynamics of the system:
the energy $C_{0}$ and all the correlations $C_{k}$ decay with
$\exp[\lambda_{1}\omega(1+\alpha)\tau]=\exp(\nu_{\HCS}^{r}t)$, see
below. On the other hand, for $\nu>\nu_c$, the dominant term is the
one corresponding to $Q_{2}\simeq 2\pi$ and the long time behaviour of
the system becomes independent of $\nu$.

The large system size limit of the eigenvalues is then
\begin{equation}
  \label{eq:lambda-large-N}  \lambda_{n}=-\frac{{Q_{n}^{(0)}}^{2}}{L^{2}}\left[1+L^{-1}\frac{2Q_{n}^{(1)}}{Q_{n}^{(0)}}+O(L^{-2})\right].
\end{equation}
Moreover, the exponent in~\eqref{eigensum} controlling the time
dependence of the contribution for each mode is
\begin{eqnarray}
  \label{eq:eigensum-exponent}
  \lambda_{n}(1+\alpha)\omega\tau&\sim&                                        
                                      -2{Q_{n}^{(0)}}^{2}\left[1+L^{-1}\frac{2Q_{n}^{(1)}}{Q_{n}^{(0)}}+O(L^{-2})\right]t,
\end{eqnarray}
which shows the relevance of the hydrodynamic scale $t$ in the large
system size limit.

\subsection{Long time limit}

Equation \eqref{eigensum} gives  the general time evolution for the
velocity correlations. Here, we show that these correlations tend to
their HCS values in the long time limit, provided that $\nu<\nu_{c}$,
that is, we are below the instability. 

Let us consider the scaled correlations $\tilde{C}_{k}$
\begin{equation}\label{eq:scaled-corr-tau}
\tilde{C}_k(\tau)=\frac{C_k(\tau)}{C_0(\tau)}=\frac{c_k(\tau)}{\sqrt{2} c_0(\tau)}, 
\end{equation}
i.e., we scale the correlations with the energy $C_{0}\neq 0$. For
long enough times, the only relevant contribution to~\eqref{eigensum}
stems from the maximum (minimum in absolute value) eigenvalue
$\lambda_1$. Thus, the time dependence for all the correlations
$C_{k}$ (or $c_{k}$) are the same and, consequently, the quotient in
\eqref{eq:scaled-corr-tau} becomes time-independent for long enough
times. Making use of \eqref{eigensum} and \eqref{eq:phi0-conv},
\begin{equation}\label{eq:scaled-corr-discrete}
\tilde{C}_k= \frac{\phi^{(1)}_k}{\sqrt{2}\phi^{(1)}_0}=-
\sin \left( \frac{q_1}{2} \right) \csc \left(
\frac{L-1}{2} q_1 \right) \cos \left[ \left( \frac{L}{2} -k \right) q_1 \right] ,
\end{equation}
which is nothing but the discrete version of \eqref{eq:stcorr1}.

We can also derive the rate at which the energy and all the
correlations are decaying in the long time limit. Particularising \eqref{eq:eigensum-exponent} for $n=1$, we have that
\begin{equation}\label{min-eigenvalue}
\lambda_1 (1+\alpha)\omega\tau \sim -\nu t\left[1-L^{-1} \psi_{\HCS}+O(L^{-2})\right] =-\nu_{\HCS}^{r}t,
\end{equation}
where $\nu_{\HCS}^{r}$ is the ``renormalised'' by fluctuations cooling
rate introduced in \eqref{eq:haff-renorm} after a multiple scale
analysis of the finite size corrections to the hydrodynamic
description. The energy is given by
$C_{0}(t)=T(t=0)\exp(-\nu_{\HCS}^{r}t)$ and the correlations
$C_{k}$ follow from~\eqref{eq:scaled-corr-discrete}.

\section{Total energy fluctuations and multiscaling} 
\label{sec:total_energy}

\begin{figure}
\centering
\includegraphics[width=0.7\textwidth]{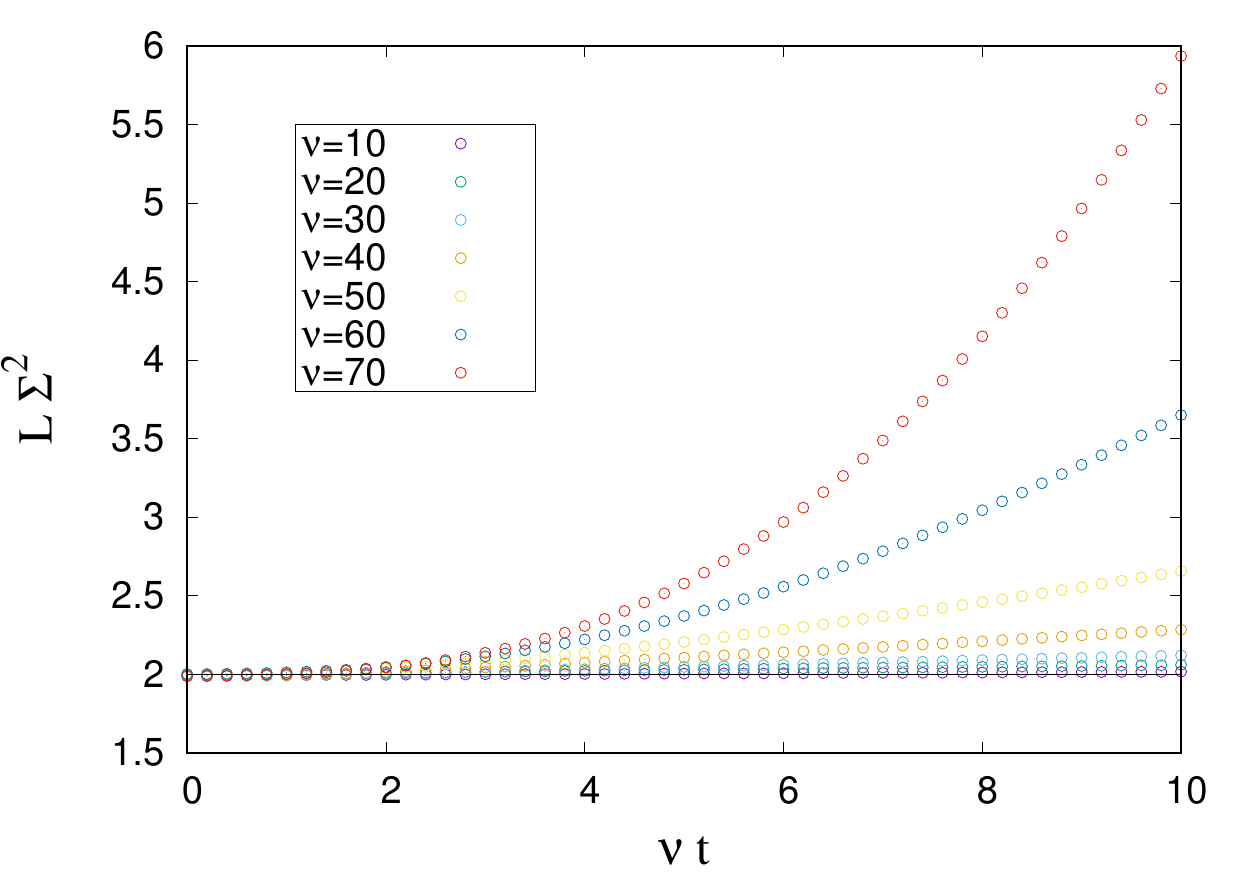}
\caption{\small Total energy rescaled fluctuations as a function of
  dimensionless time $\nu t$. We have plotted the simulation curves
  for $\nu=10,20,\ldots,70$, always with $L=1000$. The divergence from
  the expected LEA value is evident and grows with
  $\nu$.
}
\label{f:sigma_num}
\end{figure}

A typical question in granular systems concerns the distribution of
the extensive energy ${\cal K}(\tau) = \sum_l v^2_l (\tau)$: usually,
granular models present non-Gaussian distributions that can be mostly
characterized by the study of its fluctuations~\cite{BDGyM06}. Within
the same spirit of section~\ref{sec:beyond_mol_chaos_linear}, we now
aim to derive the total energy rescaled fluctuations $\Sigma(\tau)$
defined as
\begin{equation}
\label{eq:sigmadef}
\Sigma^2 (\tau) = \frac{\llangle {\cal K}^2(\tau) \rrangle - \llangle
  {\cal K}(\tau) \rrangle^2}{\llangle {\cal K}(\tau) \rrangle^2} .
\end{equation}
The Local Equilibrium Approximation (LEA) gives the
straightforward result $\Sigma^2 (\tau) = 2/L$.  However, numerical
results in figure~\ref{f:sigma_num} show a time-dependent behaviour of
$\Sigma^2(\tau)$ which clearly diverges from the LEA prediction.

Such anomalous behaviour is generally considered an evidence of
multiscaling in the moments~\cite{bnk2000}, i.e. the moments are not
scaling proportionally to the granular temperature
$T(\tau) = \llangle v^2 (\tau) \rrangle$. Notwithstanding,
  this phenomenon can also be explained by a well-defined scaled
  distribution function with some divergent moments
  \cite{Balbet,CMyP07}. Following the same approach of
section~\ref{sec:beyond_mol_chaos_linear}, we look for a direct
calculation of the energy fluctuations by means of the evolution
equations for the $4$-th order moments and correlations.

In the homogeneous case, we can write
\begin{subequations}
\label{eq:energies}
\begin{eqnarray}
\llangle {\cal K}^2 (\tau) \rrangle &=& \sum^L_{l=1} \llangle v^4_l (\tau) \rrangle + 
\sum^L_{l=1} \sum^{L-1}_{k=1} \llangle v^2_l (\tau) \, v^2_{l+k} (\tau) \rrangle \nonumber \\
&=& L \llangle v^4(\tau) \rrangle + L \left[ (L-1) T^2 (\tau) + 2 \sum^{(L-1)/2}_{k=1} 
C^{2,2}_k (\tau) \right] , \label{eq:energies-1}\\
\hspace{3mm} \llangle {\cal K} (\tau) \rrangle &=& L T(\tau) , \label{eq:energies-2}
\end{eqnarray}
\end{subequations}
where we have defined the two-particle squared velocity correlation
function
\begin{equation}
\label{eq:C22}
C^{2,2}_k (\tau) = \llangle v^2_l (\tau) v^2_{l+k} (\tau) \rrangle -
T^2(\tau), \quad k \neq 0.
\end{equation}
 Therefore, the energy fluctuations dynamics is given by the dynamics of 
$T(\tau)$, $q(t)=\llangle v^4(t) \rrangle$ and $C^{2,2}_k (\tau)$ altogether.

Rescaled energy fluctuations hence read
\begin{equation}
\label{eq:sigma2}
L \Sigma^2 (t) = 2 + a_{2}(t) + \frac{2}{T^2(t)} \int_0^{\frac{1}{2}-\frac{3}{2L}} D^{2,2} (x,t) dx ,
\end{equation}
using the hydrodynamic scaling defined in~\eqref{eq:hydro-scales} and
introducing the excess kurtosis field $a_{2}(t) = q(t)/T^2(t) -3$.
Analogously with the scaling used in
section~\ref{sec:beyond_mol_chaos}, we define $D^{2,2} = L C^{2,2}$,
where the evolution equations for these fields can be computed from
the microscopic dynamics~\eqref{eq:mom}. By means of a perturbative
approach, a set of equations is derived, similar to
\eqref{eq:0-order}, coupling all the one-particle and two-particle
fourth-degree fields, namely $q(t)$, $D^{2,2}(x,t)$ and
$D^{3,1}(x,t)$. The latter is the large-size limit of
$D^{3,1}_k (\tau) = L \llangle v^3_l (\tau) \, v_{l+k}(\tau)
\rrangle$.
Also, the three-particle correlation
$C^{1,2,1}_{i,j} (\tau) = \llangle v_{l-i} (\tau) \, v^2_l (\tau) \,
v_{l+j} (\tau) \rrangle$
appears in these equations: to get a closed set, we make use of the
\textit{clustering ansatz}, that is, we perform a cluster expansion of
the latter and neglect purely correlated terms, specifically
\begin{eqnarray}
\label{eq:cluster-exp}
C^{1,2,1}_{i,j} (\tau) &=& \llangle v^2_l (\tau) \rrangle \llangle v_{l-i} (\tau) \, v_{l+j} 
(\tau) \rrangle  + 2 \llangle v_l (\tau) \, v_{l-i} (\tau) \rrangle \llangle v_l (\tau) \, v_{l+j} 
(\tau) \rrangle  \nonumber \\
 && + \mathcal{O}(L^{-3})=\frac{1}{L} T(\tau) D_{\vert i+j \vert} (\tau) + \frac{2}{L^2} D_i(\tau) D_j(\tau) 
 + \mathcal{O}(L^{-3}). \nonumber\\
\end{eqnarray}

Using the microscopic dynamics defined in~\eqref{eq:mom} and moving to
the continuum limit defined in section~\ref{sec:beyond_mol_chaos},
with the clustering ansatz one gets to the lowest order
\begin{subequations}\label{eq:0-order-fluct}
\begin{eqnarray} 
  &&\frac{d}{dt}\tilde{q}_0(t)=0,   \label{eq:0-order-fluct-1}\\
&& \nu\left[\tilde{q}_0(t) + 3\tilde{T}^2_0\right]+8\xder{\widetilde{D}^{3,1}_0}|_{x=0}=0, \label{eq:0-order-fluct-2} \\
  &&\tder{\widetilde{D}^{3,1}_0}=\frac{\nu}{2} \left( \widetilde{D}^{3,1}_0+\tilde{T}_0\widetilde{D}_0\right)+2\,\partial_{xx}
     \widetilde{D}^{3,1}_0, \label{eq:0-order-fluct-3}  \\
  && \xder{\widetilde{D}^{3,1}_0}|_{x=1/2}=0, \label{eq:0-order-fluct-4} \\
  && \xder{\widetilde{D}^{2,2}_0}|_{x=0}=0, \label{eq:0-order-fluct-5} \\
  && \tder{\widetilde{D}^{2,2}_0}=2\,\partial_{xx} \widetilde{D}^{2,2}_0, \label{eq:0-order-fluct-6}  \\
  && \xder{\widetilde{D}^{2,2}_0}|_{x=1/2}=0. \label{eq:0-order-fluct-7}
\end{eqnarray}
\end{subequations}
These equations can be readily solved. Assuming for instance
the initial distribution to be Gaussian, we have at any time
$\tilde{q}_0=3\,\tilde{T}^2_0$. Moreover, in the long time
  limit $t \gg 1$, we obtain the stationary fields
\begin{equation}\label{eq:0-order-fluctsol}
\widetilde{D}^{3,1}_0(x) = 3 \, \widetilde{D}_0(x), \qquad
  \widetilde{D}^{2,2}_0(x)=0,
\end{equation}
recalling that $\tilde{T}_0=1$.  However, these results
do not give rise to any multiscaling effect such as the one observed
into the simulations. 

In light of the above, we move on to compute the next perturbative order. The equations needed from the
definition~\eqref{eq:sigma2} are those for $\tilde{q}_1$ and
$\widetilde{D}^{2,2}_1$, i.e.
\begin{subequations}\label{eq:1-order-fluct}
\begin{eqnarray} 
  &&\frac{d}{dt}\tilde{q}_1(t)=2\nu\psi^{3,1}_0(t),   \label{eq:1-order-fluct-1}\\
  && \nu \tilde{T}_0 \tilde{\psi}_0+\xder{\widetilde{D}^{2,2}_1}|_{x=0}=0, \label{eq:1-order-fluct-2} \\
  && \tder{\widetilde{D}^{2,2}_1}=2\,\partial_{xx} \widetilde{D}^{2,2}_1+4\,\nu\tilde{D}^2_0, \label{eq:1-order-fluct-3}  \\
  && \xder{\widetilde{D}^{2,2}_1}|_{x=1/2}=0. \label{eq:1-order-fluct-4}
\end{eqnarray}
\end{subequations}
Equation~\eqref{eq:1-order-fluct-1} is immediately solvable for long times since 
$\tilde{D}^{3,1}_0(x,t)$ is known from~\eqref{eq:0-order-fluctsol}, yielding
\begin{equation}
\label{eq:q1tilde}
\tilde{q}_1 (t) = 2 \nu \psi^{3,1}_0 = 6 \nu \psi{\HCS}.
\end{equation}
Looking at the $\tilde{D}^{2,2}_1$ field from~\eqref{eq:sigma2}, all we need is to compute the integral
$\Delta_1(t)=\int^1_0 dx \, \tilde{D}^{2,2}_1 (x,t)$. Taking
  into account~\eqref{eq:1-order-fluct}, we have that
\begin{equation}
\label{eq:d1}
\frac{d}{dt} \Delta_1 (t) = 4\nu \left[ \psi_0 (t) + \int^1_0 dx \, \widetilde{D}^2_0(x,t) \right] ,
\end{equation}
where we have used the boundary condition
$\xder{\widetilde{D}^{2,2}_1}|_{x=1^-} = -
\xder{\widetilde{D}^{2,2}_1}|_{x=0^+}$.
Therefore, in the long time limit we use the stationary
correlation profile $\widetilde{D}(x)$ from~\eqref{eq:stcorr1} to
get the stationary growth
\begin{equation}
\label{eq:d1-sol}
\frac{d}{dt} \Delta_1 (t) = 2 \nu \left[ \frac{\pi \sqrt{\nu/\nu_c}}{\sin\left(\pi \sqrt{\nu/\nu_c}\right)} \right]^2 \left[ 
1 - \frac{\sin\left(2\pi \sqrt{\nu/\nu_c}\right)}{2\pi \sqrt{\nu/\nu_c}} \right] .
\end{equation}
Now, we have all the ingredients to compute the energy fluctuations
in~\eqref{eq:sigma2}. To the first order, \eqref{eq:q1tilde}
and~\eqref{eq:T1} yield that the excess kurtosis $a_{2}(t)$
  vanishes for all times if it did initially, $a_{2}(t) = O(L^{-2})$.
  This implies that the steady-state linear divergence of the energy
fluctuations (to the first order) is given by the $D^{2,2}$
correlations term in~\eqref{eq:d1-sol}. Specifically, for
$t \gg 1$, we have
\begin{equation}
\label{eq:mnu}
\frac{d}{dt} \Sigma^2 (t) = \frac{1}{L} \frac{d}{dt} \Delta_1 (t) = \frac{\nu}{L} m_\Sigma(\nu).
\end{equation}
We have introduced $m_\Sigma(\nu) = d\Delta_1/d(\nu t)$, which is
  the slope of the energy fluctuations as a function of the
  dimensionless time $\nu t$. 

\begin{figure}[!h]
\centering
\includegraphics[width=0.7\textwidth]{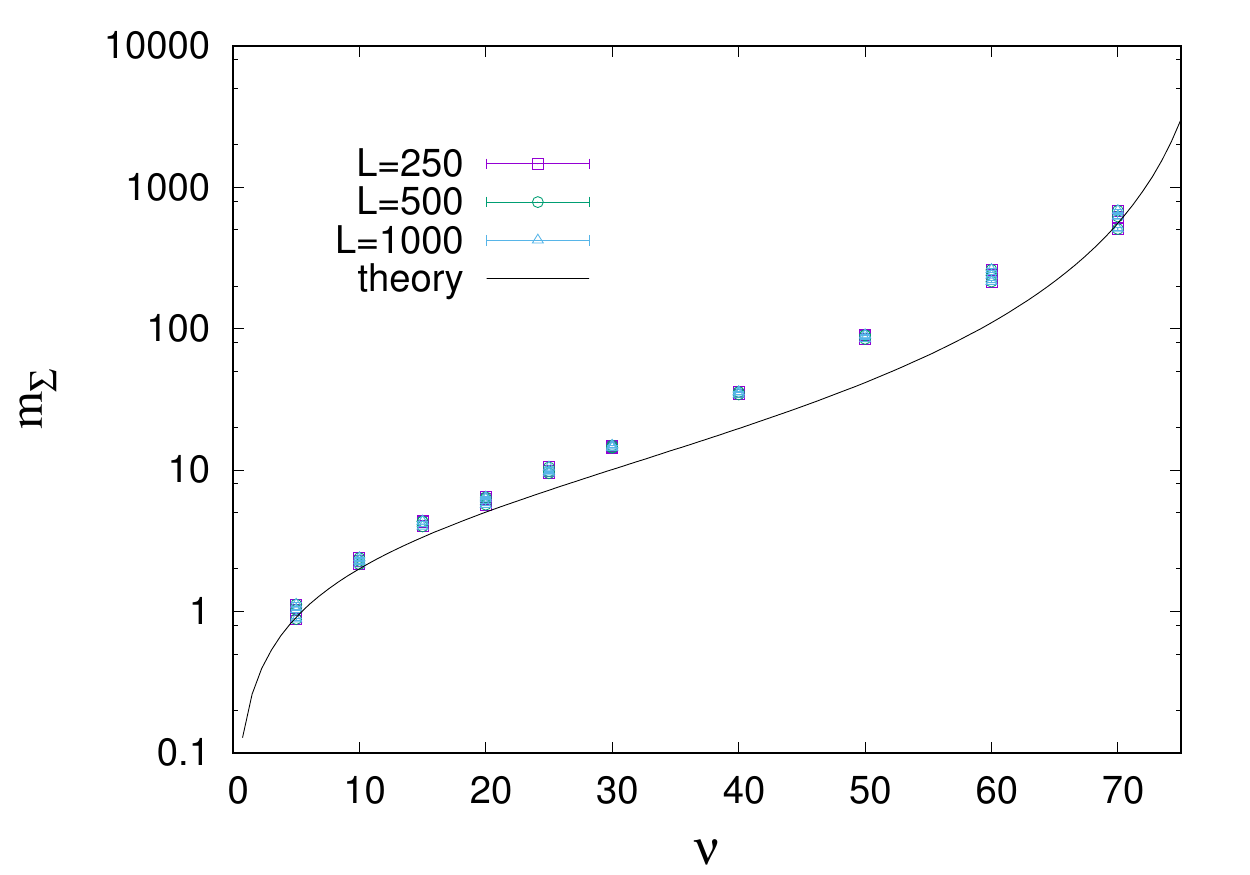}
\caption{\small Slope of the energy fluctuations vs.~time curves. We
  compare the fitted values (symbols) in the second part (long time)
  of the trajectories for $L=250,500,1000$ with the theoretical
  prediction $m_\Sigma(\nu)$ in~\eqref{eq:mnu} (black line). Simulations are
  carried out as in figure~\ref{f:sigma_num}. 
}
\label{f:energyfluct}
\end{figure}

In conclusion, the observed energy multiscaling seems to
stem from the multiscaling of many-particle correlation
fields, while the single-particle fourth moment
still scales with the granular temperature squared. We have
  compared this theoretical result with simulations in
figure~\ref{f:energyfluct}. Although some discrepancies
are apparent, especially for high $\nu$, we see that they
both exhibit a similar trend over three decades of $m_\Sigma$
  values.
Keeping the clustering ansatz in \eqref{eq:cluster-exp}, a
multiple scale analysis has also been performed, analogous
to that in
section~\ref{sec:beyond_mol_chaos}. Nevertheless, it does not
  improve the agreement with the numerics. Therefore, it seems that
  the most probable source for this discrepancy is the clustering
  ansatz that is used in both cases.

\section{Conclusions}
\label{sec:concl}

We have followed different approaches to obtain the long-range
correlations in a 1d lattice model for the velocity and energy fields
of a granular gas. The most compact approach takes advantage of a
fluctuating hydrodynamic description: it reproduces the fluctuations
and correlations of the shear-modes already known in the homogeneous
cooling of inelastic hard-spheres \cite{Noijeernst}. In
addition, it gives an approximate value of the finite-size correction
to the Haff-law describing the decay of the energy.

Fluctuating hydrodynamics however involves a continuum limit that
implies the appearance of irregular fields, a problem which is known
in the literature~\cite{Bertini2015}. On the one hand, the results of
fluctuating hydrodynamics can be reproduced and improved through the
study of correlations before taking the continuum
limit. On the other hand, still within this framework, a multiple
scale analysis makes it possible to derive the renormalisation of the
cooling rate in a more consistent way.

An exact treatment of the hierarchy of equations for the
  two-particle correlations gives access to the eigenvalue spectrum
for their time evolution. Of course, this approach
reproduces the continuum limit results when a large system is
  considered, and it is useful to understand how the
continuum limit emerges and the different timescales
that are relevant in the system. Both the energy
$C_{0}=\langle v_{l}^{2}\rangle$ and all the velocity correlations
$C_{k}=\langle v_{l}v_{l+k}\rangle$ tend to zero in the long time
limit as a consequence of the cooling, but the scaled correlations
$C_{k}/C_{0}$ become time-independent. Therefore, the system ends up
in the HCS, independently of the initial preparation.  

The above study also improves our understanding of the shear
instability in the HCS, as it comes about as a
crossing between the first two exact eigenvalues of the problem. Also,
it improves our understanding of the situation beyond the instability
($\nu>\nu_{c}$) in the present model: we find that both the energy and
the correlations still decay to zero but with a rate that is
independent of the inelasticity. Note that our system cannot
develop inhomogeneities in the density because particles are fixed.

In addition, we have also observed numerically an unexpected
multiscaling phenomenon at finite size.  While one-particle moments
scale with their corresponding power of the granular
temperature, some multiple-particle moments do
not. This implies that the variance of the total energy
  departs from its ``simple'' scaled value as time increases, with an
  approximate constant slope that seems to diverge close to the shear
  instability. We have developed a theoretical approach, based on a
  clustering hypothesis for three-particle moments, that predicts this
  multiscaling behaviour. However, the agreement between the theory
  and the simulation is not perfect, probably as a consequence of the
  clustering approximation. 

An interesting future challenge is to adapt the framework of the
Macroscopic Fluctuation Theory~\cite{Bertini,Bertini2015} to our
model. In its present state, the theory does not deal with macroscopic
equations with advection terms and momentum conservation, such as
those in the Navier-Stokes equations that inevitably appear in
granular hydrodynamics. Our model, which enforces momentum
conservation but neglects advection, represents a possible bridge
toward this goal.

\ack
  C.~A.~P.~acknowledges the support from the FPU
  Fellowship Programme of Spanish Ministerio de Educaci\'on, Cultura y
  Deporte through grant FPU14/00241. C.~A.~P.~and A.~P.~ acknowledge the
  support of the Spanish Ministerio de Econom\'{\i}a y Competitividad
  through grant FIS2014-53808-P.

\appendix

\section{Fluctuating expression for the dissipation}\label{app-a}

Let us consider the dissipation $d_{l,p}$ at site $l$ and at time $p$.
Its main part is obtained by averaging \eqref{micro-ener-dis} with
respect to the fast variables $y_{l,p}$, i.e.
\begin{equation}
  \label{eq:d-main}
  \bar{d}_{l,p}= \frac{\alpha^2-1}{4L}\left(\Delta_{l,p}^2+\Delta_{l-1,p}^2 \right)<0.
\end{equation}
This is the expression that we have to analyse in the fluctuating
hydrodynamic description, since the amplitude of the dissipation
noise scales as $L^{-3}$.  If we consider the average
  of the dissipation field, it is readily obtained that
  $d^{\av}_{l,p}=(\alpha^{2}-1)T_{l,p}/L$, which gives
  \eqref{eq:av-d-nu} in the continuum limit by using
  $d(x,t)=L^{3}d_{l,p}$ and the definition of $\nu$. Therefore, it is
  consistent to write at the fluctuating level that
\begin{equation}
  \label{eq:d-fluct}  
\bar{d}_{l,p}=\frac{\alpha^{2}-1}{L} \theta_{l,p},
\end{equation}
by defining the fluctuating temperature as
\begin{equation}\label{eq:temp-fluct}
\theta_{l,p}=\frac{1}{4}\left(\Delta_{l,p}^2+\Delta_{l-1,p}^{2}\right)=\frac{v_{l-1,p}^{2}+2v_{l,p}^{2}+v_{l+1,p}^{2}}{4}-v_{l,p}\frac{v_{l+1,p}+v_{l-1,p}}{2}.
\end{equation}
The first term on the rhs,
$(v_{l-1,p}^{2}+2v_{lp}^{2}+v_{l+1,p}^{2})/4$, reduces to $e_{l,p}$
plus terms of the order of $L^{-2}$, which are neglected.

Our main goal in to obtain a correct expression for $v_{l,p}v_{l\pm1,p}$ at
the fluctuating level. In general, we have for the average correlations
\begin{equation}
  \label{eq:discrete-corr}
  \langle v_{l,p}
  v_{l',p}\rangle=E_{l,p}\delta_{ll'}+C_{l,l'-l;p}\left(1-\delta_{ll'} \right),
\end{equation}
with the definition 
\begin{equation}
  \label{eq:def-C-dos-l}
  C_{l,l'-l;p}=\langle v_{l,p} v_{l',p}\rangle \quad \text{for $l'\neq
  l$}.
\end{equation}
The functions $C_{k,p}$ defined in section~\ref{sec:beyond_mol_chaos}
are the particularisation of $C_{l,l'-l;p}$ to an homogeneous
situation ($k=l'-l$). Consistently with \eqref{eq:discrete-corr}, we
write
\begin{equation}\label{eq:def-gamma-dos-l}
v_{l,p}v_{l',p}=e_{l,p}\delta_{ll'}+\gamma_{l,l'-l,p}\left(1-\delta_{ll'}\right)=\gamma_{l,l'-l,p}+\left(e_{l,p}-\gamma_{l,l'-l,p}\right)\delta_{ll'},
\end{equation}
at the fluctuating level. We have introduced the fluctuating
correlations $\gamma_{l,l'-l,p}$, such that
$\langle\gamma_{l,l'-l;p}\rangle=C_{l,l'-l,;p}$. In the continuum
limit, $x=l/L$ and $x'=l'/L$ and~\eqref{eq:def-gamma-dos-l} is transformed into
\begin{equation}\label{eq:def-gamma-continuum}
v(x,t)v(x',t)=\gamma(x,x'-x;t)+L^{-1}\delta(x-x')\left[e(x,t)-\gamma(x,x'-x;t)\right],
\end{equation}
because $\delta_{l,l'}\sim L^{-1} \, \delta(x-x')$ (see note at the
end of the appendix). 

Taking into account~\eqref{eq:temp-fluct} and
the above definitions, the fluctuating temperature in the continuum
limit is
\begin{equation}
\theta(x,t)=e(x,t)-\gamma(x,0;t),
\end{equation}
where we have neglected terms of the order of $L^{-2}$. Since we are interested in the limit of $\gamma(x,\Delta x;t)$ when
$\Delta x\to 0$, we use~\eqref{eq:def-gamma-continuum} with
$\Delta x=x'-x=\pm L^{-1}$ to obtain
\begin{equation}
  \label{eq:v2-continuum}
  v^{2}(x,t)=\gamma(x,0;t)+L^{-1}\left[ e(x,t)-\gamma(x,0;t)
  \right]\lim_{x'\to x}\delta(x'-x).
\end{equation}
 Thus, we have that
\begin{equation}
  \label{eq:gamma-x-0-def}
 \gamma(x,0;t)=v^{2}(x,t)-L^{-1}\theta(x,t)
  \lim_{x'\to x}\delta(x'-x).
\end{equation}
Note that $v^{2}(x,t)$ always has a singular part that stems from the
$\delta(\Delta x)$ factors on the rhs
of~\eqref{eq:v2-continuum}. Therefore, $\gamma(x,0;t)$ can be
considered as the ``regular'' part of $v^{2}(x,t)$, and  we
introduce the notation
\begin{equation}
  \label{eq:vR2-def-app}
  v_{R}^{2}(x,t)\equiv \gamma(x,0;t)=v^{2}(x,t)-L^{-1}\theta(x,t)
  \lim_{x'\to x}\delta(x'-x).
\end{equation}

By combining the previous results, and recalling that
$d(x,t)=L^{3}d_{l,p}$, we finally conclude
\begin{equation}
  \label{eq:d-cont}
  d(x,t)=-\nu\theta(x,t), \quad \theta(x,t)= e(x,t)-v_{R}^{2}(x,t).
\end{equation}
This tells us that the fluctuations of the dissipation field are
enslaved to those of the temperature. Moreover, the appearance of
$v_{R}^{2}$ in \eqref{eq:d-cont} is easy to understand on a physical
basis, since
$\langle v_{R}^{2}(x,t)\rangle=\langle\gamma(x,0;t)\rangle=
u^{2}(x,t)+O(L^{-1})$.
Equations~\eqref{eq:vR2-def-app} and \eqref{eq:d-cont} make it
possible to write a closed expression for the fluctuating temperature,
  \begin{equation}
    \label{eq:theta-closed}
    \theta(x,t)=\beta \left[e(x,t)-v^{2}(x,t)\right], \quad \beta=\left[1-L^{-1}
  \lim_{x'\to x}\delta(x'-x)\right]^{-1},
  \end{equation}
in which $\beta$ is a regularisation factor, which ``heals'' the
singularity of $v^{2}(x,t)$ in the large system size limit.

\underline{Note:} The appearance of $\delta(0)$ (more accurately,
$\lim_{x'\to x}\delta(x'-x)$) can be avoided in the
following way: for discrete $(l,l')$ we may write 
\begin{equation*}
  \delta_{ll'}=\Theta(l-l'+1/2)\Theta(l'-l+1/2),
\end{equation*}
in which $\Theta(x)$ is the Heaviside step function. Therefore, in the
continuum limit,  we have that
\begin{equation*}
  \delta_{ll'}\sim\Theta\left(x-x'+\frac{1}{2L}\right)\Theta\left(x'-x+\frac{1}{2L}\right).
\end{equation*}
When used inside an integral, the relative error introduced by using
the expression above is of the order of $L^{-2}$, since
\begin{eqnarray*}
\hspace{70mm}   \sum_{l=1}^{L} f_{l} \, \delta_{ll'}&=&f_{l'},       \\
  L \int_{0}^{1}dx f(x) \,\Theta\left(x-x'+\frac{1}{2L}\right)\Theta\left(x'-x+\frac{1}{2L}\right)&=&L \int_{x'-\frac{1}{2L}}^{{x'+\frac{1}{2L}}}dx \,
  f(x) \\
  &=& f(x')+O(L^{-2}).
\end{eqnarray*}
Therefore, both expressions, (i)
$L^{-1}\delta(x-x')$ and (ii) the product of Heaviside functions, can
be used indistinctly within the mesoscopic fluctuation framework.

Consistently with the above discussion, the Fourier components of the
product of Heaviside functions are the same as those of
$L^{.1}\delta(x-x')$, with a relative error of the order of $L^{-2}$,
\begin{eqnarray*}
  \int_{0}^{1} dx\, \Theta\left(x-x'+\frac{1}{2L}\right)\Theta\left(x'-x+\frac{1}{2L}\right) e^{-i k_{n}x}&=&\int_{x'-\frac{1}{2L}}^{{x'+\frac{1}{2L}}}dx
  e^{-ik_{n}x}\\
  &=&L^{-1}e^{-i k_{n}x'}+O(L^{-3}).
\end{eqnarray*}
Therefore,
\begin{eqnarray*}  \Theta\left(x-x'+\frac{1}{2L}\right)\Theta\left(x'-x+\frac{1}{2L}\right)&=&L^{-1}\sum_{n}e^{i
    k_{n} (x-x')}+O(L^{-3})\\
&=& L^{-1}\delta(x-x')+O(L^{-3}).
\end{eqnarray*}

\section{Sum rule up to $O(L^{-1})$}\label{app-b}

Here, we rigorously derive, in the continuum limit and up to
$O(L^{-1})$, the sum rule \eqref{eq:sum-rule-continuum} that stems
from momentum conservation.

Our starting point is
\eqref{eq:sum-rule}, which is equivalent to
\begin{equation}
  \label{eq:sum-rule-bis}
  T(t)+2\sum_{k=1}^{\frac{L-1}{2}} D_{k}(t) \Delta x=0, \quad \Delta x=L^{-1}.
\end{equation}
Now, we go to the continuum limit by making use of
\eqref{eq:hydro-scales}. To be precise, we denote here $x=(k-1)/L$ by
$x_{k}$. Then,
\begin{equation}
  \int_{x_{k}}^{x_{k+1}} dx \,D(x,t)= L^{-1} D(x_{k},t)
  +\frac{L^{-2}}{2} \partial_{x}D(x,t)|_{x_{k}}+O(L^{-3}).
\end{equation}
Hence,
\begin{eqnarray}
 \sum_{k=1}^{\frac{L-1}{2}} \underbrace{D(x_{k},t)}_{D_{k}(t)} \Delta x
 =
\int_{0}^{\frac{1}{2}-\frac{1}{2L}} dx\, D(x,t)
-\frac{L^{-1}}{2}\int_{0}^{\frac{1}{2}-\frac{1}{2L}} dx\, \partial_{x}
D(x,t)+O(L^{-2})  \nonumber \\
 = \int_{0}^{\frac{1}{2}-\frac{1}{2L}} dx\, D(x,t)
-\frac{L^{-1}}{2}\left[ D\left(\frac{1}{2}-\frac{1}{2L},t\right)-D(0,t) \right]
+O(L^{-2}).  
\end{eqnarray}
The expression above can be further simplified to
\begin{equation}\label{eq:sum-rule-intermediate}
  \sum_{k=1}^{\frac{L-1}{2}} D_{k}(t)
  \Delta x=\int_{0}^{\frac{1}{2}} dx\, D(x,t)+\frac{L^{-1}}{2}\left[
    \psi(t)-2 \chi(t) \right]+O(L^{-2}),
\end{equation}
where we have made use of the definitions of $\psi$ and $\chi$
in~\eqref{eq:psi}. If we insert \eqref{eq:sum-rule-intermediate} into
\eqref{eq:sum-rule-bis}, we obtain \eqref{eq:sum-rule-continuum} of
the main text.

\newpage

\bibliographystyle{iopart-num}
\bibliography{bibliografiav1}   

%
%

\end{document}